\documentclass[sigconf, nonacm]{acmart}

\usepackage[linesnumbered,ruled,vlined,algo2e]{algorithm2e}
\usepackage[utf8]{inputenc}
\usepackage{graphicx}
\usepackage{hyperref}
\usepackage{tabularx}
\usepackage{xcolor}
\usepackage{multirow}
\usepackage{multicol}

\usepackage{titlesec}

\usepackage{caption}
\usepackage{subcaption}
\usepackage{float}

\theoremstyle{definition}
\newtheorem{ex}{Example}

\newtheorem{definition}{Definition}

\usepackage{enumitem}

\newcommand\vldbdoi{XX.XX/XXX.XX}
\newcommand\vldbpages{XXX-XXX}
\newcommand\vldbvolume{14}
\newcommand\vldbissue{1}
\newcommand\vldbyear{2020}
\newcommand\vldbauthors{\authors}
\newcommand\vldbtitle{\shorttitle} 
\newcommand\vldbavailabilityurl{URL_TO_YOUR_ARTIFACTS}
\newcommand\vldbpagestyle{plain} 

\newcommand\hcolor{black}

\begin{document}
\title{Generalized Supervised Meta-blocking \\ \emph{\color{gray}(Technical Report)}}

\author{Luca Gagliardelli}
\affiliation{%
	\institution{Università degli studi di Modena e Reggio Emilia}
	\state{Italy}
}
\email{luca.gagliardelli@unimore.it}

\author{George Papadakis}
\affiliation{%
  \institution{National and Kapodistrian University of Athens}
  \institution{Public Power Company}
  \state{Greece}
}
\email{gpapadis@di.uoa.gr}

\author{Giovanni Simonini}
\affiliation{%
  \institution{Università degli studi di Modena e Reggio Emilia}
  \state{Italy}
}
\email{giovanni.simonini@unimore.it}

\author{Sonia Bergamaschi}
\affiliation{%
  \institution{Università degli studi di Modena e Reggio Emilia}
  \state{Italy}
}
\email{sonia.bergamaschi@unimore.it}

\author{Themis Palpanas}
\affiliation{%
  \institution{Universite  de  Paris  and  the  French  UniversityInstitute (IUF)}
  \state{France}
}
\email{themis@mi.parisdescartes.fr}

\begin{abstract}
Entity Resolution constitutes a core data integration task that relies on Blocking in order to tame its quadratic time complexity. Schema-agnostic blocking achieves very high recall, requires no domain knowledge and applies to data of any structuredness and schema heterogeneity. This comes at the cost of many irrelevant candidate pairs (i.e., comparisons), which can be significantly reduced through Meta-blocking techniques, i.e., techniques that leverage the co-occurrence patterns of entities inside the blocks: first, a weighting scheme assigns a score to every pair of candidate entities in proportion to the likelihood that they are matching and then, a pruning algorithm discards the pairs with the lowest scores. Supervised Meta-blocking goes beyond this approach by combining multiple scores per comparison into a feature vector  that is fed to a binary classifier. By using probabilistic classifiers, Generalized Supervised Meta-blocking associates every pair of candidates with a score that can be used by any pruning algorithm. For higher effectiveness, new weighting schemes are examined as features. Through an extensive experimental analysis, we identify the best pruning algorithms, their optimal sets of features as well as the minimum possible size of the training set. The resulting approaches achieve excellent performance across several established benchmark datasets.
\end{abstract}

\maketitle

\pagestyle{\vldbpagestyle}
\begingroup\small\noindent\raggedright\textbf{PVLDB Reference Format:}\\
\vldbauthors. \vldbtitle. PVLDB, \vldbvolume(\vldbissue): \vldbpages, \vldbyear.\\
\href{https://doi.org/\vldbdoi}{doi:\vldbdoi}
\endgroup
\begingroup
\renewcommand\thefootnote{}\footnote{\noindent
This work is licensed under the Creative Commons BY-NC-ND 4.0 International License. Visit \url{https://creativecommons.org/licenses/by-nc-nd/4.0/} to view a copy of this license. For any use beyond those covered by this license, obtain permission by emailing \href{mailto:info@vldb.org}{info@vldb.org}. Copyright is held by the owner/author(s). Publication rights licensed to the VLDB Endowment. \\
\raggedright Proceedings of the VLDB Endowment, Vol. \vldbvolume, No. \vldbissue\ %
ISSN 2150-8097. \\
\href{https://doi.org/\vldbdoi}{doi:\vldbdoi} \\
}\addtocounter{footnote}{-1}\endgroup

\ifdefempty{\vldbavailabilityurl}{}{
\vspace{.3cm}
\begingroup\small\noindent\raggedright\textbf{PVLDB Artifact Availability:}\\
The source code and data
have been made available at \url{https://sourceforge.net/p/erframework}.
\endgroup
}

\section{Introduction}

Entity Resolution (ER) is the task of identifying entities that describe the same real-world object among different datasets~\cite{DBLP:journals/csur/ChristophidesEP21}. 
ER constitutes a core data integration task with many applications that range from Data Cleaning in databases to Link Discovery in Semantic Web data~\cite{DBLP:series/synthesis/2015Christophides,DBLP:series/synthesis/2015Dong}. Despite the bulk of works on ER, it remains a challenging task \cite{DBLP:journals/csur/ChristophidesEP21}. 
One of the main reasons is its quadratic time complexity: in the worst case, every entity has to be compared with all others, thus scaling poorly 
to large volumes of data.

To tame its high complexity, \textit{Blocking} is typically used \cite{DBLP:journals/tkde/Christen12,DBLP:journals/csur/PapadakisSTP20}. 
Instead of considering all possible pairs of entities, it restricts the computational cost to entities that are similar. 
This is efficiently carried out by associating every entity with one or more signatures and clustering together entities that have identical or similar signatures. 
Extensive experimental analyses have demonstrated that the \textit{schema-agnostic} signatures outperform the schema-based ones, without requiring domain or schema knowledge \cite{papadakis2012blocking,DBLP:series/synthesis/2021Papadakis}. 
As a result, parts of any attribute value in each entity can be used as signatures.


\begin{figure}
	\centering
	\includegraphics[width=\linewidth]{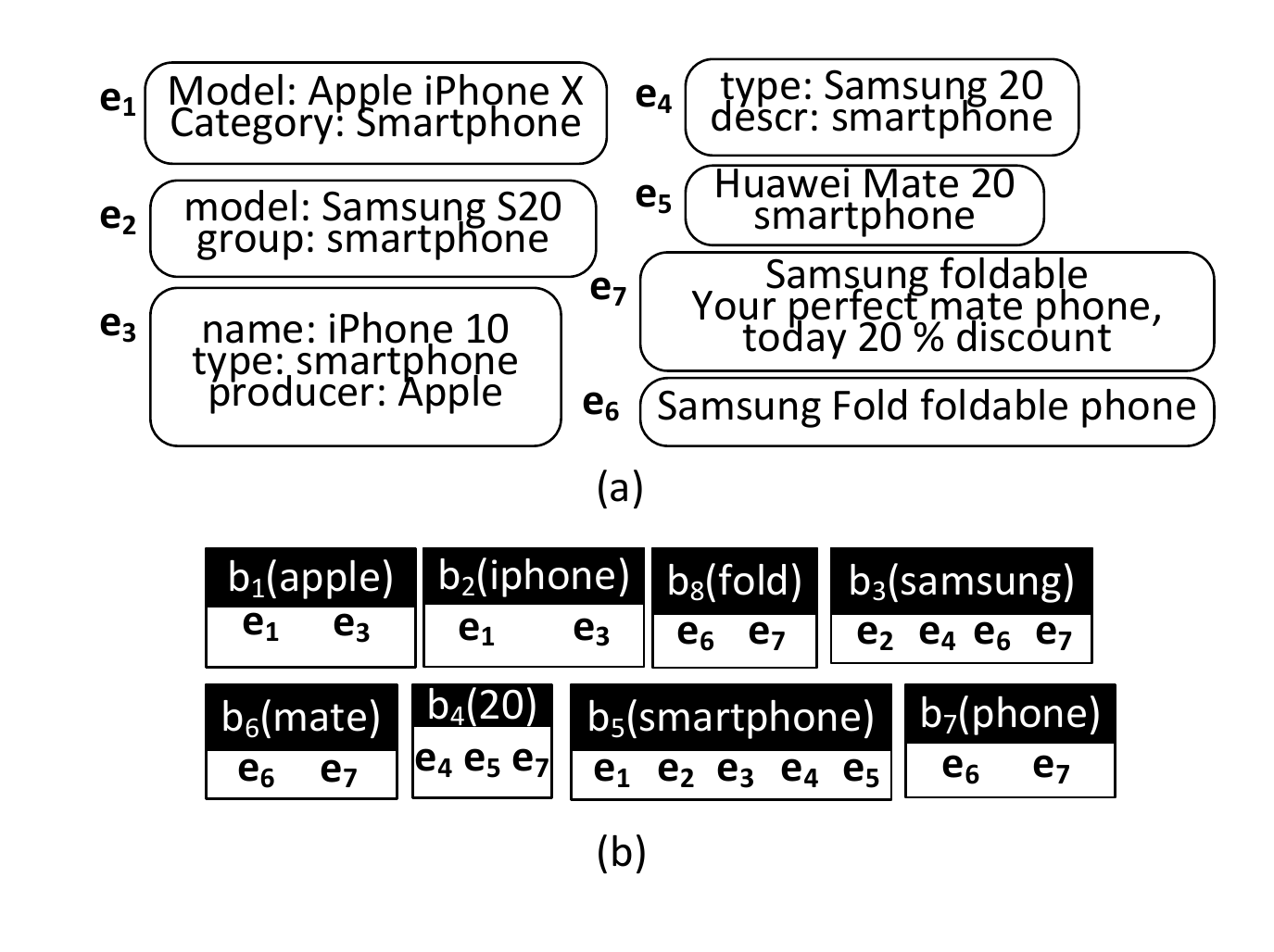}
	\vspace{-25pt}
	\caption{(a) The input entities (smartphone models), and (b) the redundancy-positive blocks produced by Token Blocking.}
	\label{fig:blocking}
	\vspace{-10pt}
\end{figure}

\begin{ex}[Schema-agnostic blocking]
	\it
An example of schema-agnostic blocking is illustrated in Figure \ref{fig:blocking}.
The profiles in Figure \ref{fig:blocking}a contain {\color{\hcolor}three duplicate pairs $\langle e_1$, $e_3 \rangle$, $\langle e_2$, $e_4 \rangle$ and $\langle e_6$, $e_7 \rangle$.}
The profiles are clustered together by using Token Blocking, i.e., a block is created for every token that appears in the values of each profile.
ER examines all pairs inside each block and, thus, can detect all duplicate pairs, as they co-occur in at least one block.
\end{ex}

On the downside, the resulting blocks involve high levels of redundancy: every entity is associated with multiple blocks, thus yielding numerous \textit{redundant} and \textit{superfluous comparisons} \cite{DBLP:journals/jdiq/BeneventanoBGS20,DBLP:journals/pvldb/SimoniniBJ16}. The former are pairs of entities that are repeated across different blocks, while the latter involve non-matching entities. For example, the pair $\langle e_1$, $e_3 \rangle$ is redundant in $b_2$, as it is already examined in $b_1$, while the pair $\langle e_2$, $e_6 \rangle \in b_3$ is superfluous, as the two entities are not duplicates. Both types of comparisons can be skipped, reducing the computational cost of ER without any impact on recall \cite{DBLP:journals/pvldb/0001APK15,DBLP:journals/pvldb/0001SGP16}.

A core approach to this end is \textit{Meta-blocking} \cite{DBLP:journals/tkde/PapadakisKPN14}, which discards all redundant comparisons, while reducing significantly the portion of superfluous ones. It relies on two components to achieve this goal:
\begin{enumerate}[leftmargin=*]
    \item A \textit{weighting scheme} is a function that receives as input a pair of entities along with their associated blocks and returns a score proportional to their matching likelihood.
    The score is based on the co-occurrence patterns of the entities into the original set of blocks: the more blocks they share and the more distinctive (i.e., infrequent) the corresponding signatures are, the more likely they are to match and the higher is their score.
    \item A \textit{pruning algorithm} receives as input all weighted pairs and retains the ones that are more likely to be matching.
\end{enumerate}

\begin{figure}[]
	\begin{subfigure}[b]{0.4\linewidth}
		\centering
		\includegraphics[width=\linewidth]{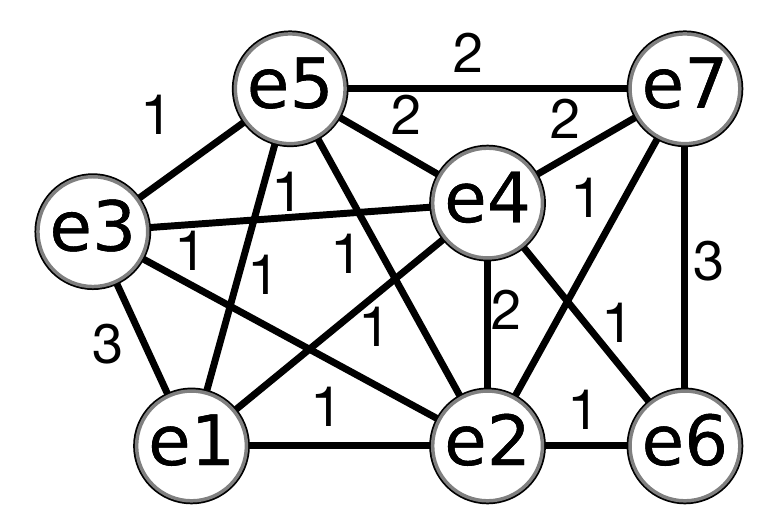}
		\vspace{-15pt}
		\caption{}
		\label{fig:toy:es_01_a}
	\end{subfigure}
	\hspace{0pt}
	\begin{subfigure}[b]{0.33\linewidth}
		\centering
		\includegraphics[width=\linewidth]{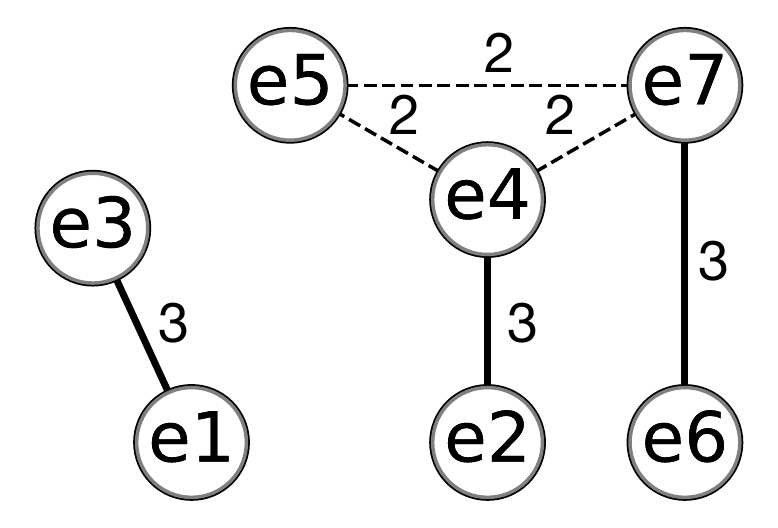}
		\vspace{-15pt}
		\caption{}
		\label{fig:toy:es_01_b}
	\end{subfigure}
	\hspace{0pt}
	\begin{subfigure}[b]{0.23\linewidth}
		\centering
		\includegraphics[width=\linewidth]{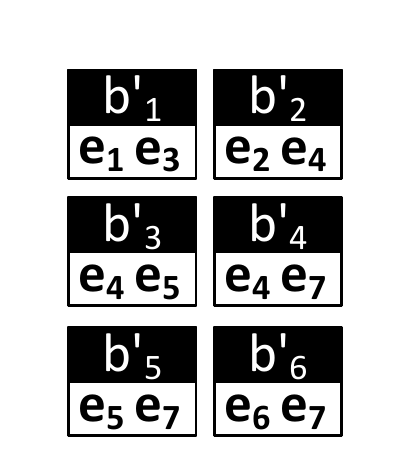}
		\vspace{-15pt}
		\caption{}
		\label{fig:toy:es_01_c}
	\end{subfigure}
	\vspace{-10pt}
	\caption{Unsupervised Meta-blocking example: (a) The blocking graph of the blocks in Figure \ref{fig:blocking}b, using the number of common blocks as edge weights, (b) a possible pruned blocking graph, and (c) the new blocks.}
	\label{fig:unsupervised_mb}
	\vspace{-10pt}
\end{figure}

\begin{ex}[Unsupervised Meta-blocking]
	\it
	Unsupervised Meta-blocking builds a 
	blocking graph (Figure \ref{fig:toy:es_01_a}) from the blocks in Figure \ref{fig:blocking}b as follows: each entity profile is represented as a node;
	two nodes are connected by an edge if the corresponding profiles co-occur in at least one block;
	each edge is weighted according to a weighting scheme---in our example, the number of blocks shared by the adjacent profiles.
	Finally, the blocking graph is pruned according to a pruning algorithm---in our example, for each node, we discard the edges with a weight lower than the average of its edges.
	The pruned blocking graph appears in Figure \ref{fig:toy:es_01_b}, with the dashed lines representing the superfluous comparisons.
	A new block is then created for every retained edge. Figure \ref{fig:blocking}c presents the final blocks, which involve
	significantly fewer pairs without missing the matching ones.
	This is a schema-agnostic process, just like the original blocking method.
\end{ex}

\begin{figure}[]
	\begin{subfigure}[b]{0.4\linewidth}
		\centering
		\includegraphics[width=\linewidth]{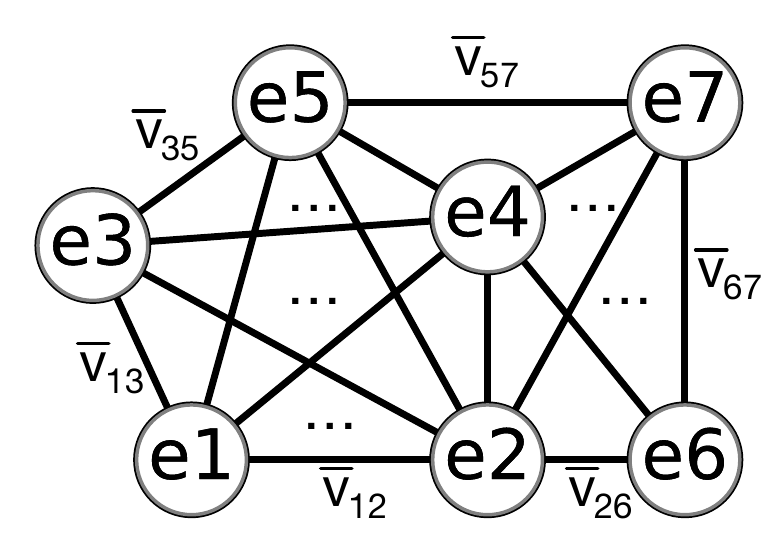}
		\vspace{-15pt}
		\caption{}
		\label{fig:toy:es_02_a}
	\end{subfigure}
	\hspace{0pt}
	\begin{subfigure}[b]{0.33\linewidth}
		\centering
		\includegraphics[width=\linewidth]{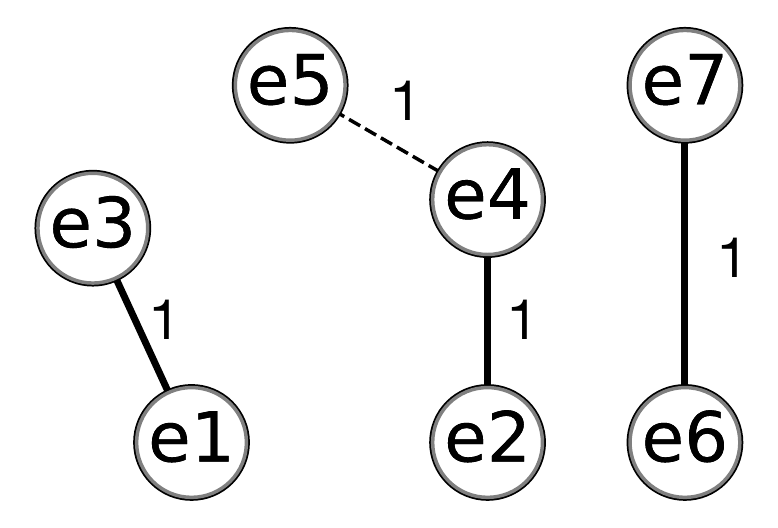}
		\vspace{-15pt}
		\caption{}
		\label{fig:toy:es_02_b}
	\end{subfigure}
	\hspace{0pt}
	\begin{subfigure}[b]{0.22\linewidth}
		\centering
		\includegraphics[width=\linewidth]{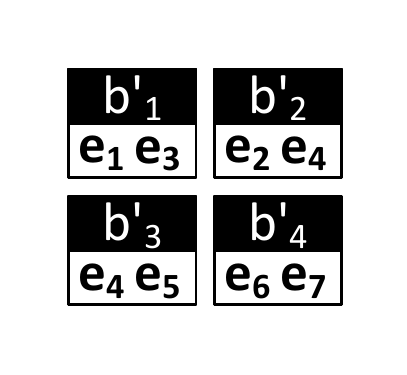}
		\vspace{-15pt}
		\caption{}
		\label{fig:toy:es_02_c}
	\end{subfigure}
	\vspace{-10pt}
	\caption{Supervised Meta-blocking example: (a) a graph where each edge is associated with a feature vector; (b) the graph pruned by employing a binary classifier (c) the output contains a new block per remaining edge.}
	\label{fig:supervised_mb}
	\vspace{-10pt}
\end{figure}

\subsection{Supervised Meta-blocking}

Supervised Meta-blocking models the restructuring of a set of blocks as a binary classification task~ \cite{papadakis2014supervised}. Its goal is to train a model that learns to classify every comparison as \textit{positive} (i.e., likely to be matching) and \textit{negative} (i.e., unlikely to be matching). To this end, every pair is associated with a feature vector that comprises a combination of the most distinctive weighting schemes that are used by unsupervised meta-blocking. Thus, Supervised Meta-blocking considers more comprehensive evidence, outperforming the unsupervised approaches to a significant extent.

In more detail, the thorough experimental analysis in~\cite{papadakis2014supervised} performed an analytical feature selection that considered all combinations of 7 features to determine the one achieving the best balance between effectiveness and efficiency. The resulting vector comprises four features,
yielding high time efficiency and classification accuracy.

\begin{ex}[Supervised Meta-blocking]
	\it
	Figure \ref{fig:toy:es_02_a} shows
	the feature vectors generated for every distinct comparison {\color{\hcolor}(i.e., edge in the blocking graph)} in the blocks of Figure \ref{fig:blocking}b. 
	For instance, each pair of entities $\langle e_i, e_j \rangle$ can be represented by a feature vector {\color{\hcolor} $v_{i,j}=\{CB(e_i, e_j), JS(e_i, e_j)\}$}, where $CB(e_i, e_j)$ is the number of their common blocks and $JS(e_i, e_j)$ is the Jaccard coefficient of blocks associated with $e_i$ and $e_j$.
	Then, a \underline{binary classifier} is trained with a sample of labelled vectors and is used to predict whether a pair  $\langle e_i, e_j \rangle$ is a match ($l_{i,j}$=1) or not ($l_{i,j}$=0). The results appear in Figure \ref{fig:toy:es_02_b}, {\color{\hcolor} where the dashed lines represent superfluous comparisons.} The comparisons classified as positive are retained, yielding the new blocks
	in Figure~\ref{fig:toy:es_02_c}.
\end{ex}

Note, though, that ER suffers from intense class imbalance, since the vast majority of comparisons belongs to the negative class. To address it, \textit{undersampling} is used to create a training set that is equally split between the two classes. Through extensive experiments, the number of labelled instances per class in the training set was set to 5\% of the minority (positive) class in the ground-truth \cite{papadakis2014supervised}. This size combined high effectiveness with high efficiency, through the learning of generic classification models. Additionally, the learned classification models were empirically verified to be robust with respect to the configuration of the classification algorithm: minor changes in the internal parameters of the algorithm yield minor changes in its overall performance.

Even though Supervised Meta-blocking outperforms its unsupervised counterpart to a significant extent, there is plenty of room for improving its classification accuracy, for reducing its overhead, i.e., running time, and for minimizing the size of the training set it requires, as we explain below.

\begin{figure}[]
	\begin{subfigure}[b]{0.4\linewidth}
		\centering
		\includegraphics[width=\linewidth]{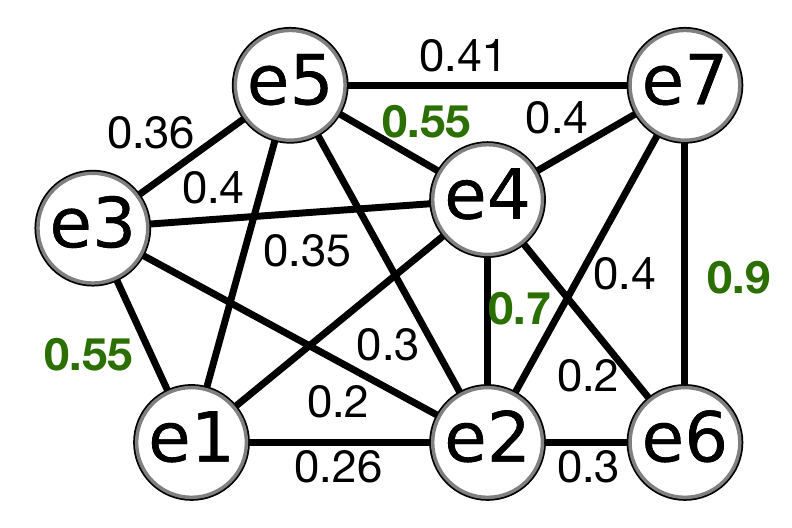}
		\vspace{-15pt}
		\caption{}
		\label{fig:toy:es_03_a}
	\end{subfigure}
	\hspace{0pt}
	\begin{subfigure}[b]{0.33\linewidth}
		\centering
		\includegraphics[width=\linewidth]{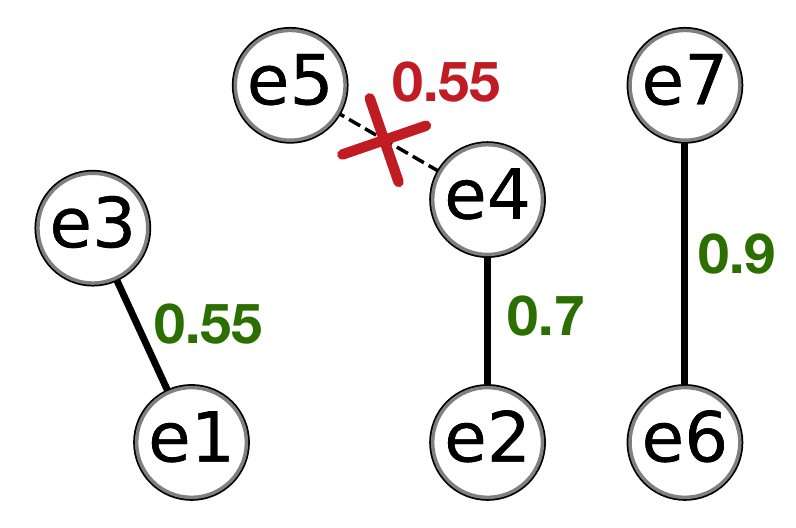}
		\vspace{-15pt}
		\caption{}
		\label{fig:toy:es_03_b}
	\end{subfigure}
	\hspace{0pt}
	\begin{subfigure}[b]{0.15\linewidth}
		\centering
		\includegraphics[width=\linewidth]{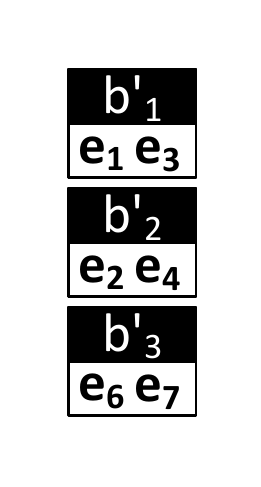}
		\vspace{-15pt}
		\caption{}
		\label{fig:toy:es_03_c}
	\end{subfigure}
	\vspace{-10pt}
	\caption{
		Generalized Supervised Meta-blocking example:
		(a) a graph weighted with a probabilistic classifier; (b) the pruned graph; and (c) the new blocks.}
	\label{fig:gen_supervised_mb}
	\vspace{-10pt}
\end{figure}

{\color{\hcolor}
	Supervised Meta-blocking requires the effort to provide labeled edges, but by representing each edge with multiple features, it is more accurate in
	discriminating matching and non-matching pairs than Unsupervised Meta-blocking, which 
	employs a single weight per edge.
	Indeed, Supervised Meta-blocking consistently yields better precision and recall than the unsupervised approach~\cite{papadakis2014supervised}.
	Yet, 
	the binary classifier that lies at its core acts as
	a learned, unique, \emph{global} threshold that is then employed to prune the edges. 
	Defining  \emph{local} threshold 
	for each node would allow a finer control on which edges to prune.
	This is the intuition behind Generalized Supervised Meta-blocking, as illustrated in the following example.
	
	\begin{ex}[Generalized Supervised Meta-blocking]
		\it
		Our new approach builds a graph where every edge is associated with a feature vector 
		(as Supervised Meta-blocking does in Figure~\ref{fig:toy:es_02_a}) and trains a \underline{probabilistic classifier}, which assigns a weight (the matching probability) to each edge (Figure~\ref{fig:toy:es_03_a}).
		Then, several weight- and cardinality-based algorithms can be applied. For example, \textsf{Supervised WNP} it prunes the graph as follows: for each node, all its adjacent edges with a weight lower that $0.5$ are discarded; for the remaining edges, only those with a weight greater than the average on are kept. Figure~\ref{fig:toy:es_03_b} shows the result of this step: two edges may be assigned the same weight by the probabilistic classifier, e.g., $\langle e_1, e3\rangle$ and $\langle e_4, e5\rangle$, but they may be kept (e.g., the matching pair $\langle e_1, e3\rangle$) or discarded (e.g., the non-matching pair $\langle e_4, e5\rangle$) depending on their context, i.e., the weights in their neighborhood.
		Note that $\langle e4, e5 \rangle$ is not discarded by Supervised Meta-blocking in Figure \ref{fig:toy:es_02_b}, which thus underperforms
		Supervised WNP in terms of precision (for the same recall).
	\end{ex}
}

\subsection{Our Contributions}

{\color{\hcolor}Our work is motivated by a real-world application that aims to deduplicate a legacy customer database. It contains $\sim$7.5 million entries that correspond to electricity supplies and, thus, are associated with an address, a customer name as well as a series of optional information (e.g., tax id) that are typically empty. To exploit all available information, the computational cost is reduced through schema-agnostic blocking. Our goal is actually to minimize the set of candidate pairs, using Supervised Meta-blocking, while restricting human involvement for the generation of the labelled instances. To this end, }
we go beyond Supervised Meta-blocking in the following ways:

\begin{enumerate}[leftmargin=*]
    \item We generalize it from a binary classification task to a binary \textit{probabilistic} classification process. The resulting  probabilities are used as comparison weights, on top of which we apply new pruning algorithms that are incompatible with the original approach.
    \item Using the original features, we prove that the new pruning algorithms significantly outperform the existing ones through an extensive experimental study that involves 9 real-world datasets. We also specify the top performers among the weight- and cardinality-based algorithms.
    \item To further improve their performance, we use four new weighting schemes 
    as features for Generalized Supervised Meta-blocking. 
    We 
    examine the performance of all 255 feature combinations over all 9 datasets for the top-performing algorithms. 
    We identify the top-10 feature sets per algorithm in terms of effectiveness and then, we select the optimal one by considering their time efficiency, too, significantly reducing the overall run-time.
    \item For each algorithm, we examine how the size of the training set affects its performance.
    We experimentally demonstrate that it suffices to train our approaches over 
    just 50 labelled instances, 25 from each class.
    \item 
    We performn a scalability analysis over 5 synthetic datasets with up to 300,000 entities, 
    proving that our approaches scale well both with respect to effectiveness and time-efficiency under versatile settings.
    \item We have publicly released our data as well as an implementation of all pruning algorithms and weighting schemes in Java.\footnote{See \url{https://sourceforge.net/p/erframework} for more details.}
\end{enumerate}

The rest of the paper is organized as follows: Section \ref{sec:preliminaries} provides background knowledge on the task of Supervised Meta-blocking, Section \ref{sec:pruningAlgorithms} introduces the new pruning algorithms, and Section \ref{sec:weightingSchemes} discusses the weighing schemes that are used as features. The experimental analysis is presented in Section \ref{sec:experiments}, the main works in the field are discussed in Section \ref{sec:relatedWork} and the paper concludes in Section \ref{sec:conclusions} along with directions for future work.  
\section{Preliminaries}
\label{sec:preliminaries}

An entity profile $e_i$ is defined as a set of name-value pairs, i.e., $e_i = \{\langle n_j, v_j\rangle\}$, where both the attribute names and the attribute values are textual. This simple model is flexible and generic enough to seamlessly accommodate a broad range of established data formats -- from the structured records in relational databases to the semi-structured entity descriptions in RDF data \cite{papadakis2012blocking}. Two entities, $e_i$ and $e_j$, that describe the same real-world object are called \textit{duplicates} or \textit{matches}, denoted by $e_i \equiv e_j$. A set of entities is called \textit{entity collection} and is denoted by $E_l$. An entity collection $E_l$ is \textit{clean} if it is duplicate-free, i.e., $\not \exists~ e_i, e_j \in E_l : e_i \equiv e_j$.

In this context, we distinguish Entity Resolution into two tasks \cite{DBLP:journals/tkde/Christen12,DBLP:series/synthesis/2021Papadakis}: (i) \textit{Clean-Clean ER} or \textit{Record Linkage} receives as input two clean entity collections, $E_1$ \& $E_2$, and detects the set of duplicates $D$ between their entities, $D = \{(e_i, e_j) \subseteq E_1 \times E_2 : e_i \equiv e_j\}$; (ii)
\textit{Dirty ER} or \textit{Deduplication} receives as input a dirty entity collection and detects the duplicates it contains, $D = \{(e_i, e_j) \subseteq E \times E : i \neq j \wedge e_i \equiv e_j\}$.

In both cases, the time complexity is quadratic with respect to the input, i.e., $O(|E_1|\times|E_2|)$ and $O(|E|^2)$, respectively, as every entity profile has to be compared with all possible matches. To reduce this high computational cost, Blocking restricts the search space to similar entities~\cite{DBLP:journals/csur/PapadakisSTP20}.

Meta-blocking operates on top of BLocking, refining an existing set of blocks, $B$, a.k.a. \textit{block collection}, as long as it is \textit{redundancy-positive}. This means that every entity $e_i$ participates into multiple blocks, i.e., $|B_i|\geq 1$, where $B_i=\{b \in B : e_i \in b\}$ denotes the set of blocks containing $e_i$, and the more blocks two entities share, the more likely they are to be matching, because they share a larger portion of their content. Blocks of this type are generated by methods like Token Blocking, Suffix Arrays and Q-Grams Blocking and their variants~\cite{DBLP:journals/pvldb/0001APK15,DBLP:journals/pvldb/0001SGP16}.

The redundancy-positive block collections involve a large portion of \textit{redundant comparisons}, as the same pairs of entities are repeated across different blocks. These can be easily removed by aggregating for every entity $e_i \in E_1$ the \underline{set} of all entities from $E_2$ that share at least one block with it \cite{papadakis2016scaling}. The union of these individual sets yields the distinct set of comparisons, which is called \textit{candidate pairs} and is denoted by $C$. Every non-redundant comparison between $e_i$ and $e_j$, $c_{i,j} \in C$, belongs to one of the following~types:
\begin{itemize}
    \item \textit{Positive pair} if $e_i$ and $e_j$ are matching: $e_i \equiv e_j$.
    \item \textit{Negative pair} if $e_i$ and $e_j$ are not matching: $e_i \not\equiv e_j$.
\end{itemize}

Note that these definitions are independent of Matching: two matching (non-matching) entities are positive (negative) as long as they share at least one block in $B$ \cite{DBLP:journals/tkde/Christen12,papadakis2012blocking}.
The set of all positive and negative pairs in a block collection $B$ are denoted by $P_B$ and $N_B$, respectively.  The goal of Meta-blocking is to transform a given block collection $B$ into a new one $B'$ such that the negative pairs are drastically reduced without any significant impact on the positive ones, i.e., $|P_{B'}| \approx |P_B|$ and $|N_{B'}| \ll |N_B|$.


\subsection{Problem Definition}

Supervised Meta-blocking models every pair $c_{i,j} \in C$ as a feature vector $f_{i,j} = [s_1(c_{i,j}), s_2(c_{i,j}), ..., s_n(c_{i,j}) ]$, where each $s_i$ is a weighting scheme that returns a score proportional to the matching likelihood of $c_{i,j}$. The feature vectors for all pairs in $C$ are fed to a \textit{binary classifier}, which labels them as \texttt{positive} or \texttt{negative}, if their constituent entities are highly likely to match or not.
We assess the performance of this process based on the following measures:
\begin{enumerate}[leftmargin=*]
    \item $TP(C)$, the true positive pairs, involve matching entities and are correctly classified as \texttt{positive}.
    \item $FP(C)$, the false positive pairs, entail non-matching entities, but are classified as \texttt{positive}.
    \item $TN(C)$, the true negative pairs, entail non-matching entities and are correctly classified as \texttt{negative}.
    \item $FN(C)$, the false negative pairs, comprise matching entities, but are categorized as \texttt{negative}.
\end{enumerate}

Supervised Meta-blocking discards all candidate pairs labelled as \texttt{negative}, i.e., $TN(C) \cup FN(C)$, retaining those belonging to $TP(C) \cup FP(C)$. A new block is created for every \texttt{positive} pair, yielding the new block collection $B'$. Thus, the effectiveness of Supervised Meta-blocking is assessed with respect to the following measures:
\begin{itemize}[leftmargin=*]
    \item \textit{Recall}, a.k.a. \textit{Pairs Completeness}, expresses the portion of existing duplicates that are 
    retained, i.e.,\\
    $Re = |TP(C)|/|D| = (|D|-FN(C))/|D|$.
    \item \textit{Precision}, a.k.a. \textit{Pairs Quality}, expresses the portion of positive candidate pairs that are indeed matching, i.e., $Pr = |TP(C)|/ (|TP(C)|+|FP(C)|)$.
    \item \textit{F-Measure} is the harmonic mean of recall and precision, i.e., $F1= 2 \cdot Re \cdot Pr / (Re + Pr)$.
\end{itemize}

All measures are defined in $[0,1]$, with higher values indicating higher effectiveness.

In this context, the task of Supervised Meta-blocking is formally defined as follows \cite{papadakis2014supervised}:

\begin{definition}
\label{def:SM}
Given the candidate pairs $C$ of block collection $B$, the labels $L$=\{\texttt{positive}, \texttt{negative}\} and a training set $T = \{\langle c_{i,j}, l_k\rangle : c_{i,j} \in C \wedge l_k \in L\}$, Supervised Meta-blocking aims to learn a classification model $M$ that minimizes the cardinality of $FN(C) \cup FP(C)$ so that the new block collection $B'$ achieves much higher precision than $B$, $Pr(B')$$\gg$$Pr(B)$, but maintains the original
recall, $Re(B')$$\approx$$Re(B)$. 
\end{definition}



The time efficiency of Supervised Meta-blocking is assessed through its running time, $RT$. This includes the time required to: (i) generate the feature vectors for all candidate pairs in $C$, (ii) 
train the classification model $M$, and (iii) 
apply the trained classification model $M$ to $C$.

\subsubsection{Generalized Supervised Meta-blocking}

This new task differs from Supervised Meta-blocking in two ways: (i) instead of a \textit{binary} classifier that assigns class labels, it trains a \textit{probabilistic} classifier that assigns a weight $w_{i,j} \in [0,1]$ to every candidate pair $c_{i,j}$. This weight expresses how likely it is to belong to the positive class. (ii) The candidate pairs with a probability lower than 0.5 are discarded, but the rest, which are called \textit{valid pairs}, are further processed by a pruning algorithm. The ones retained after pruning give raise to the new block collection $B'$, i.e., $B'$ contains a new block per retained valid pair.

Hence, the performance evaluation of Generalized Supervised Meta-blocking relies on the following measures:
\begin{enumerate}[leftmargin=*]
    \item $TP'(C)$, the probabilistic true positive pairs, involve matching entities that are assigned a probability $\geq$0.5 and are retained by the pruning algorithm.
    \item $FP'(C)$, the probabilistic false positive pairs, entail non-matching entities, that are assigned a probability $\geq$0.5 and are retained by the pruning algorithm.
    \item $TN'(C)$, the probabilistic true negative pairs, entail non-matching entities that are assigned a probability $<$0.5 and are discarded by the pruning algorithm.
    \item $FN'(C)$, the probabilistic false negative pairs, comprise matching entities, that are assigned a probability $<$0.5 and are discarded by the pruning algorithm.
\end{enumerate}

The measures of recall, precision and F-Measure are redefined accordingly. In this context, the task of Generalized Supervised Meta-blocking is formally defined as follows:

\begin{definition}
\label{def:GSM}
Given the  candidate pairs $C$ of block collection $B$, the labels $L$=\{\texttt{positive}, \texttt{negative}\}, and a training set $T = \{\langle c_{i,j}, l_k\rangle : c_{i,j} \in C \wedge l_k \in L\}$, the goal of Generalized Supervised Meta-blocking is to train a probabilistic classification model $M$ that assigns a weight $w_{i,j} \in [0,1]$ to every candidate pair $c_{i,j} \in C$; these weights are then processed by a pruning algorithm so as to minimize the cardinality of $FN'(C) \cup FP'(C)$, yielding a new block collection $B'$ that achieves much higher precision than $B$, $Pr(B') \gg Pr(B)$, while maintaining the original
recall, $Re(B') \approx Re(B)$.
\end{definition}

The time efficiency of Generalized Supervised Meta-blocking is assessed through its run-time, $RT$, which adds to that of Supervised Meta-blocking the time required to process the assigned probabilities by a pruning algorithm.

\section{Pruning algorithms}
\label{sec:pruningAlgorithms}

A supervised pruning algorithm operates as follows: given a specific set of features, it trains a probabilistic classifier on the available labelled instances.
Then, it applies the trained classification model $M$ to each candidate pair, estimating its classification probability. It it exceeds 0.5, it is compared with a threshold in order to determine whether the corresponding pair of entities will be retained or not.

Depending on the type of threshold, the pruning algorithms are categorized into two types:
\begin{enumerate}[leftmargin=*]
    \item The \textit{weight-based algorithms} determine the weight(s), above which a comparison is retained.
    \item The \textit{cardinality-based algorithms} determine the number $k$ of top-weighted comparisons to be retained.
\end{enumerate}

In both cases, the determined threshold is applied either \textit{globally}, on all candidate pairs, or \textit{locally}, on the candidate pairs associated with every individual entity. Below, we delve into the supervised algorithms of each category.

\subsection{Weight-based pruning algorithms}

This category includes the following four algorithms. None of them was considered in \cite{papadakis2014supervised} - only \textsf{WEP} was approximated through the binary classification task in Definition~\ref{def:SM}.

\vspace{4pt}
\noindent
\textbf{Weigted Edge Pruning (\textsf{WEP}).} Algorithm \ref{algo:wep} iterates over the set of candidate pairs $C$ twice: first, it applies the trained classifier to each pair in order to estimate the average probability $\bar{p}$ of the valid ones (Lines 1-8). Then, it applies again the trained classifier to each pair and retains only those pairs with a probability higher than $\bar{p}$ (Lines 9-13).

\begin{algorithm2e}[t]
\DontPrintSemicolon
\KwIn{Learned Model $M$, Candidate Pairs $C$}
\KwOut{New Candidate Pairs $C'$}

$\bar{p}$ $\leftarrow$ 0\;
counter = 0\;
\ForEach{$c_{i,j} \in C$}{
    $p_{i,j}$ $\leftarrow$ $M$.getProbability($c_{i,j})$\;
    \If{0.5 $\leq$ $p_{i,j}$}{
        $\bar{p}$ += $p_{i,j}$ \;
        counter += 1\;
    }
}
$\bar{p}$ $\leftarrow$ $\bar{p}$ / counter\;

$C'$ $\leftarrow$ \{\}\;
\ForEach{$c_{i,j} \in C$}{
    $p_{i,j}$ $\leftarrow$ $M$.getProbability($c_{i,j})$\;
    \If{$\bar{p}$ $\leq$ $p_{i,j}$}{
        $C'$ $\leftarrow$  $C' \cup \{c_{i,j}\}$\;
    }
}
\textbf{return} $C'$
\caption{Supervised Weighted Edge Pruning}
\label{algo:wep}
\end{algorithm2e}
\begin{algorithm2e}[t]
\DontPrintSemicolon
\KwIn{Learned Model $M$, Candidate Pairs $C$}
\KwOut{New Candidate Pairs $C'$}

$\bar{p}$[] $\leftarrow$ \{\}\;
$counter$[] $\leftarrow$ \{\}\;
\ForEach{$c_{i,j} \in C$}{
    $p_{i,j}$ $\leftarrow$ $M$.getProbability($c_{i,j})$\;
    \If{0.5 $\leq$ $p_{i,j}$}{
        $\bar{p[i]}$ += $p_{i,j}$ \;
        $counter[i]$ += 1\;
        $\bar{p[j]}$ += $p_{i,j}$ \;
        $counter[j]$ += 1\;
    }
}
\ForEach{$i \in |\bar{p}|$}{
    $\bar{p[i]}$ $\leftarrow$ $\bar{p[i]}$ / $counter[i]$\;
}
$C'$ $\leftarrow$ \{\}\;
\ForEach{$c_{i,j} \in C$}{
    $p_{i,j}$ $\leftarrow$ $M$.getProbability($c_{i,j})$\;
    \If{$\bar{p[i]}$ $\leq$ $p_{i,j}$ $\vee$ $\bar{p[j]}$ $\leq$ $p_{i,j}$}{
        $C'$ $\leftarrow$  $C' \cup \{c_{i,j}\}$\;
    }
}
\textbf{return} $C'$
\caption{Supervised Weighted Node Pruning}
\label{algo:wnp}
\end{algorithm2e}

\vspace{4pt}
\noindent
\textbf{Weighted Node Pruning (\textsf{WNP}).} Algorithm \ref{algo:wnp} iterates twice over $C$, too. Yet, instead of a global average probability, it estimates a local average probability per entity. To this end, it keeps in memory two arrays: $\bar{p}$[] with the sum of valid probabilities per entity (Line 1) and $counter$[] with the number of valid candidates per entity (Line 2). These arrays are populated during the first iteration over $C$ (Lines 3-9). The average probability per entity is then computed in Lines 10-11. Finally, \textsf{WNP} iterates over $C$ and retains every comparison $c_{i,j}$ only if its estimated probability $p_{i,j}$ exceeds either of the related average probabilities (Line 15). 

\vspace{4pt}
\noindent
\textbf{Reciprocal Weighted Node Pruning (\textsf{RWNP}).} The only difference from \textsf{WNP} is that a comparison is retained if its classification probability exceeds both related average probabilities, i.e., $\bar{p[i]}$ $\leq$ $p_{i,j}$ $\mathbf{\wedge}$ $\bar{p[j]}$ $\leq$ $p_{i,j}$. This way, it applies a consistently deeper pruning than \textsf{WNP}.

\vspace{4pt}
\noindent
\textbf{BLAST.} This algorithm is similar to \textsf{WNP}, but uses a fundamentally different pruning criterion. Instead of the average probability per entity, it relies on the maximum probability per entity $e_i$. Algorithm \ref{algo:blast} stores these probabilities in the array $max[]$ (Line 1), which is populated during the first iteration over $C$ (Lines 2-8).
The second iteration over $C$ retains a valid pair $c_{i,j}$ if it exceeds a certain portion $r$ of the sum of the related maximum probabilities (Line 12).

\begin{algorithm2e}[tb]
\DontPrintSemicolon
\KwIn{Learned Model $M$, Candidate Pairs $C$, Pruning Ratio $r \in (0,1]$}
\KwOut{New Candidate Pairs $C'$}

$max$[] $\leftarrow$ \{\}\;
\ForEach{$c_{i,j} \in C$}{
    $p_{i,j}$ $\leftarrow$ $M$.getProbability($c_{i,j})$\;
    \If{0.5 $\leq$ $p_{i,j}$}{
        \If{$max[i]$ $<$ $p_{i,j}$}{
            $max[i]$ = $p_{i,j}$ \;
        }
        
        \If{$max[j]$ $<$ $p_{i,j}$}{
            $max[j]$ = $p_{i,j}$ \;
        }
    }
}

$C'$ $\leftarrow$ \{\}\;
\ForEach{$c_{i,j} \in C$}{
    $p_{i,j}$ $\leftarrow$ $M$.getProbability($c_{i,j})$\;
    \If{0.5 $\leq$ $p_{i,j}$ $\wedge$ $r \cdot (max[i]+max[j])$ $\leq$ $p_{i,j}$}{
        $C'$ $\leftarrow$  $C' \cup \{c_{i,j}\}$\;
    }
}
\textbf{return} $C'$
\caption{Supervised BLAST}
\label{algo:blast}
\end{algorithm2e}

\vspace{-10pt}
\subsection{Cardinality-based pruning algorithms}

This category includes the three algorithms described below. Only the first two were considered in \cite{papadakis2014supervised}.

\vspace{4pt}
\noindent
\textbf{Cardinality Edge Pruning (\textsf{CEP}).} This algorithm retains the top-$K$ weighted comparisons among the candidate pairs, where $K$ is set to half the sum of block sizes in the original block collection $B$, i.e., $K = \sum_{b_i \in B}|b|/2$, where $|b|$ stands for the number of entities in block $b$ \cite{DBLP:journals/tkde/PapadakisKPN14}. Algorithm \ref{algo:cep} essentially maintains a priority queue $Q$ (Line 1), which sorts the comparisons in decreasing probability. $Q$ is populated through a single iteration over $C$ (Lines 3-9). Every valid candidate pair that exceeds the minimum probability $min_p$ (Line 5), is pushed to the queue (Line 6). Whenever the size of the queue exceeds $K$, the lowest-weighted comparison is removed from the queue and $min_p$ is updated accordingly (Lines 7-9). At the end of the iteration, the contents of $Q$ correspond to the new set of candidates~$C'$.

\begin{algorithm2e}[tbh]
\DontPrintSemicolon
\KwIn{Learned Model $M$, Candidate Pairs $C$, $K$}
\KwOut{New Candidate Pairs $C'$}

$Q$ $\leftarrow$ \{\}\;
$min_p$ $\leftarrow$ 0\;
\ForEach{$c_{i,j} \in C$}{
    $p_{i,j}$ $\leftarrow$ $M$.getProbability($c_{i,j})$\;
    \If{0.5 $\leq$ $p_{i,j}$ $\wedge$ $min_p < p_{i,j}$}{
        $Q$.push($c_{i,j}$)\;
        \If{$K < |Q|$}{
            $c_{k,l}$ $\leftarrow$ $Q$.pop()\;  
            $min_p$ $\leftarrow$ $p_{k,l}$\;
        }
    }
}
\textbf{return} $Q$
\caption{Supervised Cardinality Edge Pruning}
\label{algo:cep}
\end{algorithm2e}

\noindent
\textbf{Cardinality Node Pruning (\textsf{CNP}).} Algorithm \ref{algo:cnp} adapts \textsf{CEP} to a local operation, maintaining an array $Q[]$ with a separate priority queue per entity (Line 1). The maximum size of each queue depends on the characteristics of the original block collection, as it amounts to the average number of blocks per entity: $k=max(1, \sum_{b \in B}|b|/(|E_1|+|E_2|))$ \cite{DBLP:journals/tkde/PapadakisKPN14}. During the first iteration over $C$, \textsf{CNP} populates the priority queue of every entity following the same procedure as \textsf{CEP} (Lines 3-15); if the probability of the current candidate pair exceeds the minimum probability of one of the relevant queues (Lines 6 and 11), the pair is pushed into the queue (Lines 7 and 12). Whenever the size of a queue exceeds $k$ (Lines 8 and 13), the least-weighted comparison is removed (Lines 9 and 14) and the corresponding threshold is updated accordingly (Lines 10 and 15). \textsf{CNP} involves a second iteration over $C$ (Lines 17-21), which retains a candidate pair $c_{i,j}$ if its contained in the priority queue of $e_i$ or $e_j$ (Line 20). 

\begin{algorithm2e}[tbh]
\DontPrintSemicolon
\KwIn{Learned Model $M$, Candidate Pairs $C$, $k$}
\KwOut{New Candidate Pairs $C'$}

$Q[]$ $\leftarrow$ \{\}\;
$min_p[]$ $\leftarrow$ \{\}\;
\ForEach{$c_{i,j} \in C$}{
    $p_{i,j}$ $\leftarrow$ $M$.getProbability($c_{i,j})$\;
    \If{0.5 $\leq$ $p_{i,j}$}{
        \If{$min_p[i]$ $<$ $p_{i,j}$}{
            $Q[i]$.push($c_{i,j}$)\;
            \If{$k < |Q[i]|$}{
                $c_{l,m}$ $\leftarrow$ $Q[i]$.pop()\;  
                $min_p[i]$ $\leftarrow$ $p_{l,m}$\;
            }
        }
        \If{$min_p[j]$ $<$ $p_{i,j}$}{
            $Q[j]$.push($c_{i,j}$)\;
            \If{$k < |Q[j]|$}{
                $c_{l,m}$ $\leftarrow$ $Q[j]$.pop()\;  
                $min_p[j]$ $\leftarrow$ $p_{l,m}$\;
            }
        }
    }
}
$C'$ $\leftarrow$ \{\}\;
\ForEach{$c_{i,j} \in C$}{
    $p_{i,j}$ $\leftarrow$ $M$.getProbability($c_{i,j})$\;
    \If{0.5 $\leq$ $p_{i,j}$}{
        \If{$Q[i]$.contains($c_{i,j}$) $\vee$ $Q[j]$.contains($c_{i,j}$)}{
            $C'$ $\leftarrow$  $C' \cup \{c_{i,j}\}$\;
        }
    }
}
\textbf{return} $C'$
\caption{Supervised Cardinality Node Pruning}
\label{algo:cnp}
\end{algorithm2e}

\vspace{4pt}
\noindent
\textbf{Reciprocal Cardinality Node Pruning (\textsf{RCNP}).} This algorithm adapts \textsf{CNP} so that it performs a consistently deeper pruning, requiring that every retained comparison is contained in the priority queue of both constituent entities. That is, the condition of Line 20 in Algorithm \ref{algo:cnp} changes into a conjunction: $Q[i]$.contains($c_{i,j}$) $\mathbf{\wedge}$ $Q[j]$.contains($c_{i,j}$).
\section{Weighting Schemes}
\label{sec:weightingSchemes}

The goal of \textit{weighting schemes} is to infer the matching likelihood of candidate pairs from their co-occurrence patterns in the input blocks \cite{DBLP:journals/tkde/PapadakisKPN14}.  
All schemes are schema-agnostic, being generic enough to apply to any redundancy-positive block collection. In \cite{papadakis2014supervised}, four weighting schemes formed the optimal feature vector in the sense that it achieves the best balance between effectiveness and time efficiency:


\begin{enumerate}[leftmargin=*]
    \item \textit{Co-occurrence Frequency-Inverse Block Frequency} (\textsf{CF-IBF}). Inspired from Information Retrieval's TF-IDF, it assigns high scores to entities that participate in few blocks, but co-occur in many of them. More formally:
    \[
    CF-IBF(c_{i,j})=|B_i\cap B_j| \cdot \log \frac{|B|}{|B_i|} \cdot \log \frac{|B|}{|B_j|}. 
    \]
    
    \item \textit{Reciprocal Aggregate Cardinality of Common Blocks} (\textsf{RACCB}). The smaller the blocks shared by a pair of candidates, the more distinctive information they have in common and, thus, the more likely they are to be matching. This idea is captured by the following sum:
    \[RACCB(c_{i,j})=\sum_{b \in B_i\cap B_j} \frac{1}{||b||},\]
    where $||b||$ denotes the total number of candidate pairs in block $b$ (including the redundant ones).
    \item \textit{Jaccard Scheme} (\textsf{JS}). It expresses the portion of blocks shared by a pair of candidates:
    \[JS(c_{i,j})=%
    \frac{|B_i\cap B_j|}{|B_i|+|B_j|-|B_i\cap B_j|}
    \]
    This captures the core characteristic of redundancy-positive block collections that the more blocks two entities share, the more likely they are to match.
    
    \item \textit{Local Candidate Pairs} (\textsf{LCP}).  It measures the number of candidates for a particular entity. More formally:
    \[LCP(e_i)=|\{e_j: i \neq j \wedge |B_i\cap B_j| > 0\}|.\]
    
    The rationale is that the less candidate matches correspond to an entity, the more likely it is to match with one of them. Entities with many candidates convey no distinctive information, being unlikely for any match.
\end{enumerate}

The last feature applies to an individual entity. Thus, the feature vector of 
$c_{ij}$ includes 
both \textsf{LCP}($e_i$) and \textsf{LCP}($e_j$) \cite{papadakis2014supervised}.

In this work, we aim to enhance the effectiveness of the resulting feature vector. To this end, we additionally consider the following new weighting schemes \cite{TRLIPADE}:
\begin{enumerate}[leftmargin=*]
\item \textit{Enhanced Jaccard Scheme} (\textsf{EJS}). Similar to TF-IDF, it enhances \textsf{JS} with the inverse frequency of an entity's candidates in the set of all candidate pairs:
    \[
    EJS(c_{i,j})=%
    JS(c_{i,j}) \cdot \log \frac{||B||}{||e_i||} \cdot \log \frac{||B||}{||e_j||},
    \]
    where {\footnotesize $||B||=\sum_{b \in B}||b||$} and {\footnotesize $||e_l|| = \sum_{b \in B_l}||b||$}. 
\item \textit{Weighted Jaccard Scheme} (\textsf{WJS}). Its goal is to alter \textsf{JS} so that it considers the size of the blocks containing every entity, promoting the smallest (and most distinctive) ones in terms of the total number of candidates. Thus, it multiplies every block in the Jaccard coefficient with its inverse size:
{\scriptsize
\[
WJS(c_{i,j})=
\frac{\sum_{b \in  B_i\cap B_j} \frac{1}{||b||}}
{\sum_{b \in B_i} \frac{1}{||b||} + \sum_{b \in B_j} \frac{1}{||b||} - \sum_{b \in B_i \cap B_j} \frac{1}{||b||}}.
\]
}
\textsf{WJS} can be seen as normalizing \textsf{RACCB}.
\item \textit{Reciprocal Sizes Scheme} (\textsf{RS}). It is similar to \textsf{ARCS}, but considers the number of entities in common blocks, rather than the number of candidate pairs:
\[
RS(c_{i,j})=\sum_{b \in B_i\cap B_j} \frac{1}{|b|}.
\]
\item \textit{Normalized Reciprocal Sizes Scheme} (\textsf{NRS}). It normalizes \textsf{RS}, multiplying every block in the Jaccard coefficient with its inverse size:
{\scriptsize
\[
NRS(c_{i,j})=\frac{\sum_{b \in B_i \cap B_j} \frac{1}{|b|}}{\sum_{b \in B_i} \frac{1}{|b|} + \sum_{b \in B_j} \frac{1}{|b|} - \sum_{b \in  B_i\cap B_j} \frac{1}{|b|}}.
\]
}
\end{enumerate}


\section{Experimental evaluation}
\label{sec:experiments}

\subsection{Experimental setup}

\textbf{Hardware and Software}---All the experiments are performed on a machine equipped with four Intel Xeon E5-2697 2.40 GHz (72 cores), 216 GB of RAM, running Ubuntu 18.04.
We employed the \textit{SparkER} library \cite{sparker} to perform blocking and features generation.

Unless stated otherwise, we perform machine learning analysis using Python 3.7.
In particular, all classification models are implemented through the Support Vector Classification (SVC) model of scikit-learn\footnote{\url{https://scikit-learn.org/stable/modules/generated/sklearn.svm.LinearSVC.html}}. We used the default configuration parameters, enabling the generation of probabilities and fixing the random state so as to reproduce the probabilities over several runs. Note that we have performed all experiments with logistic regression, too, obtaining almost identical results. Due to space limitations, 
we do not report them in this paper.

\begin{table}[t]
    \centering
    \footnotesize
    \begin{tabularx}{\linewidth}{|X|c|c|c|c|}
        \hline
        \textbf{Name} & \textbf{$\mathbf{|E_1|}$} & \textbf{$\mathbf{|E_2|}$} & \textbf{$\mathbf{|D}$}  & \textbf{$\mathbf{|C|}~~~$} \\ \hline\hline
        AbtBuy & 1.1k & 1.1k & 1.1k & 36.7k \\ \hline
        DblpAcm & 2.6k & 2.3k & 2.2k &  46.2k \\ \hline
        ScholarDblp & 2.5k & 61.3k & 2.3k & 83.3k \\ \hline
        AmazonGP & 1.4k & 3.3k & 1.3k & 84.4k \\ \hline
        ImdbTmdb & 5.1k & 6.0k & 1.9k & 109.4k \\ \hline
        ImdbTvdb & 5.1k & 7.8k & 1.1k & 119.1k \\ \hline
        TmdbTvdb & 6.0k & 7.8k & 1.1k & 198.6k \\ \hline
        Movies & 27.6K & 23.1k & 22.8k & 26.0M \\ \hline
        WalmartAmazon & 2.5K & 22.1k & 1.1k & 27.4M \\ \hline
	\end{tabularx}
	\vspace{0.5em}
    \caption{Technical characteristics of the real-world Clean-Clean  ER datasets used in the experiments. 
    $|E_x|$ stands for the number of entities in a constituent dataset, $|D|$ for the number of duplicate pairs, and $|C|$ for the number of distinct candidate pairs generated in the corresponding block collection. The datasets are ordered in decreasing~$|C|$.
    }
    \vspace{-25pt}
    \label{tab:datasets}
\end{table}

\vspace{4pt}
\noindent 
\textbf{Datasets}---Table \ref{tab:datasets} lists the 9 real-world datasets employed in our experiments.
They have different characteristics and cover a variety of domains.
Each dataset involves two different, but overlapping data sources, where the ground truth of the real matches is known.
\texttt{AbtBuy} matches products extracted from Abt.com and Buy.com \cite{kopcke2010evaluation}. \texttt{DblpAcm} matches scientific articles extracted from dblp.org and dl.acm.org \cite{kopcke2010evaluation}. \texttt{ScholarDblp} matches scientific articles extracted from scholar.google.com and dblp.org \cite{kopcke2010evaluation}.
\texttt{ImdbTmdb}, \texttt{ImdbTvdb} and \texttt{TmdbTvdb} match movies and TV series extracted from IMDB, TheMovieDB and TheTVDB \cite{obraczka2021eager}, as suggested by their names.
\texttt{Movies} matches information about films that are extracted from imdb.com and dbpedia.org \cite{papadakis2012blocking}.
Finally, \texttt{WalmartAmazon} matches products from Walmart.com and Amazon.com \cite{magellandata}.

\vspace{4pt}
\noindent  
\textbf{Blocking}---For each dataset, the initial block collection is extracted through Token Blocking,the only parameter-free redundancy-positive blocking method \cite{DBLP:journals/csur/PapadakisSTP20}.
The original block collection is then processed by Block Purging \cite{papadakis2012blocking}, which discards all the blocks that contain more than half of the entity profiles in the collection in a parameter-free way. These blocks correspond to highly frequent signatures (e.g. stop-words) that provide no distinguishing information. Subsequently, we apply Block Filtering \cite{papadakis2016scaling}, 
removing each entity $e_i$ from the largest 20\% blocks in which it appears.
Finally, the features described in Section \ref{sec:weightingSchemes} are generated for each pair of candidates.

Table \ref{tab:blocking_results} reports the performance of the resulting block collections.
We observe that in most cases, the block collections achieve an almost perfect recall that significantly exceeds 90\%. The only exception is \texttt{AmazonGP}, where some duplicate entities share no \textit{infrequent} attribute value token - the recall, though, remains quite satisfactory, even in this case. Yet, the precision is consistently quite low, as its highest value is lower than 0.003. As a result, F1 is also quite low, far below 0.1 across all datasets. \textit{These settings undoubtedly call for Supervised Meta-blocking to raises precision and F1 by orders of magnitude, at a small cost~in~recall.}

To apply Generalized Supervised Meta-blocking to these block collections, we take special care to avoid the bias derived from the randomly selected pairs to be labelled and used for training. To this end, we performed 10 runs and averaged the values of precision, recall, and F1. In each run, a different seed is used to sample the pairs that compose the training set. Using undersampling, we formed a balanced training set per dataset that comprises \textbf{500} labelled instances, equally split between the positive and the negative class.
Due to space limitations, in the following we mostly report the average performance of every approach across all 9 block collections.

\begin{table}[]
    \centering
    \footnotesize
    \begin{tabular}{|l|c|c|c|}
    \hline
    \textbf{Dataset}     & \textbf{Recall} & \textbf{Precision} & \textbf{$F1$} \\ \hline
    \hline
    AbtBuy               & 0.948           & 2.78$\cdot$10$^{-2}$           & 5.40$\cdot$10$^{-2}$          \\ \hline
    DblpAcm              & 0.999           & 4.81$\cdot$10$^{-2}$           & 9.18$\cdot$10$^{-2}$          \\ \hline
    ScholarDblp          & 0.998           & 2.80$\cdot$10$^{-3}$           & 5.58$\cdot$10$^{-3}$          \\ \hline
    AmazonGP & 0.840           & 1.29$\cdot$10$^{-2}$           & 2.54$\cdot$10$^{-2}$          \\ \hline
    ImdbTmdb             & 0.988           & 1.78$\cdot$10$^{-2}$           & 3.50$\cdot$10$^{-2}$          \\ \hline
    ImdbTvdb             & 0.985           & 8.90$\cdot$10$^{-3}$           & 1.76$\cdot$10$^{-2}$          \\ \hline
    TmdbTvdb             & 0.989           & 5.50$\cdot$10$^{-3}$           & 1.09$\cdot$10$^{-2}$          \\ \hline
    Movies               & 0.976           & 8.59$\cdot$10$^{-4}$           & 1.72$\cdot$10$^{-3}$          \\ \hline
    WalmartAmazon        & 1.000           & 4.22$\cdot$10$^{-5}$           & 8.44$\cdot$10$^{-5}$          \\ \hline
    \end{tabular}
    \vspace{0.5em}
    \caption{Performance of the block collections that are given as input to the supervised meta-blocking methods.}
    \label{tab:blocking_results}
    \vspace{-20pt}
\end{table}



\subsection{Pruning Algorithm Selection}
The goal of this experiment is to discover which are the best-performing weighted-based and cardinality-based pruning algorithms for Generalized Supervised Meta-blocking among those discussed in Section \ref{sec:pruningAlgorithms}. As baseline methods, we employ the pruning algorithms proposed in \cite{papadakis2014supervised}: the binary classifier that approximates \textsf{WEP} for weight-based algorithms, denoted by \textbf{\textsf{BCl}} in the following, as well as \textsf{CEP} and \textsf{CNP} for the cardinality-based ones.
We fixed the training set size to 500 pairs and used the feature vector proposed in \cite{papadakis2014supervised} as optimal; that is, every candidate pair $c_{i,j}$ is represented by the following vector:
$\{CF-IBF(c_{i,j}), RACCB(c_{i,j}), JS(c_{i,j}),\\ LCP(e_i), LCP(e_j)\}$.
Based on preliminary experiments, we set the pruning ratio of BLAST to $r$=0.35.

\begin{figure}[t]
\centering
\includegraphics[width=\linewidth]{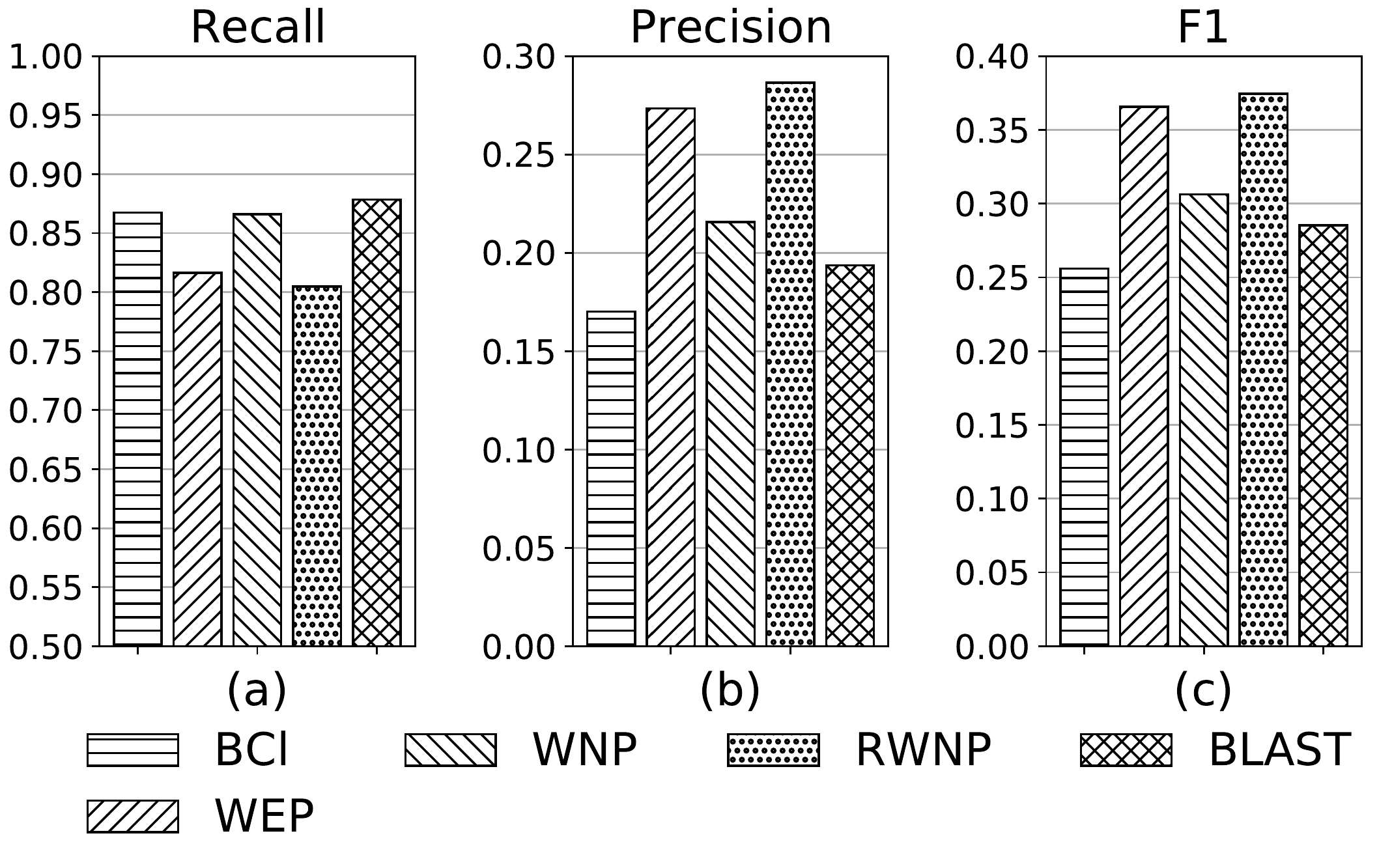}
\vspace{-15pt}
\caption{The average performance of all weight-based pruning algorithms over the block collections of Table \ref{tab:blocking_results}.}
\vspace{-10pt}
\label{fig:w_based}
\end{figure}

\begin{figure}[t]
\centering
\includegraphics[width=\linewidth]{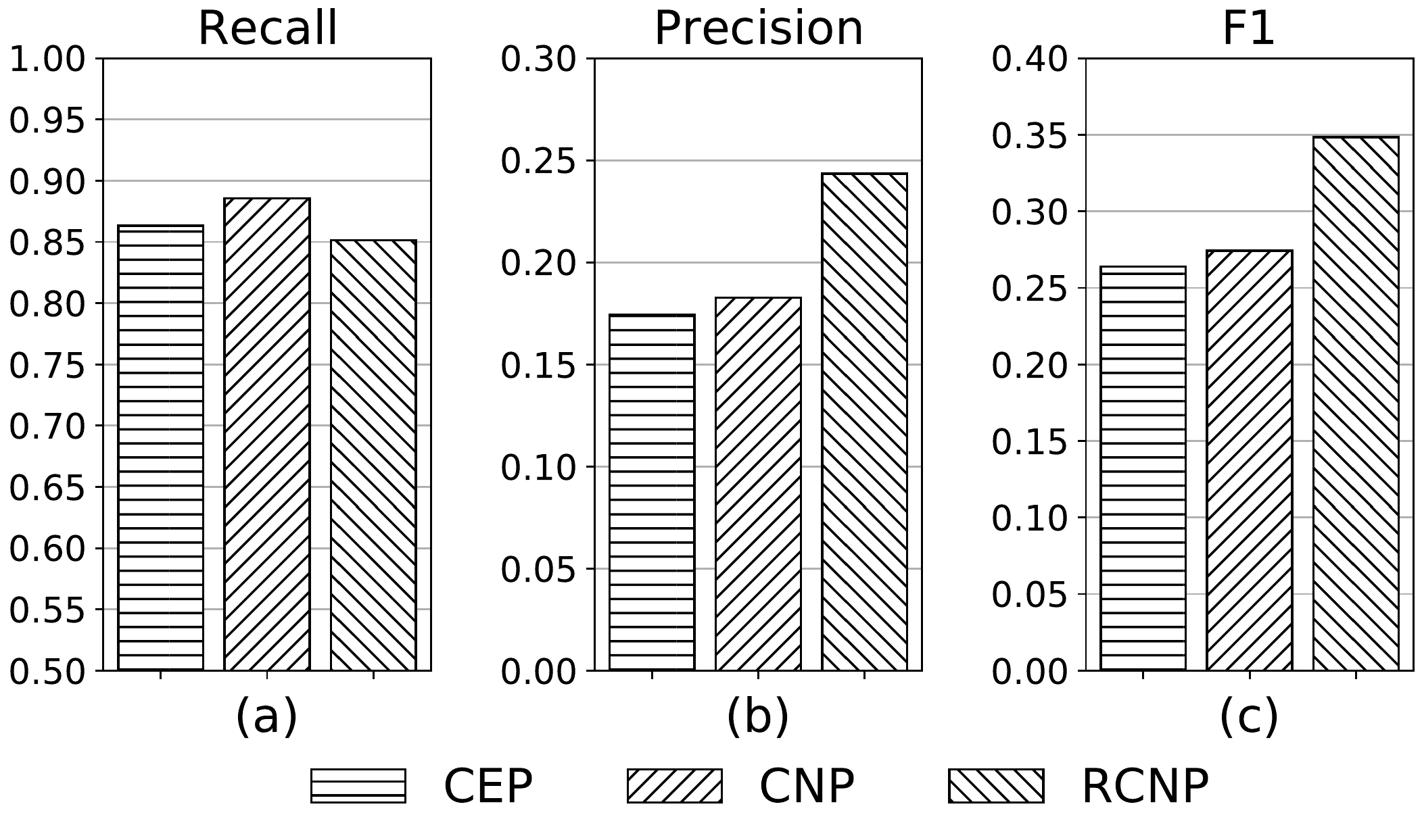}
\vspace{-15pt}
\caption{The average performance of all cardinality-based pruning algorithms over the block collections of Table \ref{tab:blocking_results}.}
\vspace{-10pt}
\label{fig:c_based}
\end{figure}

The average effectiveness measures of the weight- and cardinality
based algorithms across the 9 block collections of Table \ref{tab:blocking_results} are reported in Figures \ref{fig:w_based} and \ref{fig:c_based}, respectively.



Among the 
the weight-based algorithms, we observe that the new pruning algorithms trade sliglthy lower recall for significantly higher precision and F1. Comparing \textsf{BCl} with \textsf{WEP}, recall drops by -5.9\%, while precision raises by 60.8\% and F1 by 42.9\%. This pattern is more intense in the case of \textsf{RWNP}: it reduces recall by -7.2\%, increasing precision by 68.5\% and F1 by 46.3\%. These two algorithms actually monopolize the highest F1 scores in every case: for \texttt{ImdbTmdb}, \texttt{ImdbTvdb} and \texttt{TmdbTvdb}, \textsf{WEP} ranks first with \textsf{RWNP} second and vice versa for the rest of the datasets. Thus, \textsf{RWNP} achieves the highest average F1 (0.374), followed in close distance by \textsf{WEP} (0.366). However, their aggressive pruning results in very low recall ($\ll$0.8) in four datasets. E.g., in the case of \texttt{AbtBuy}, \textsf{BCl}'s recall is 0.852, but \textsf{WEP} and \textsf{RWNP} reduce it to 0.755 and 0.699, resp.

The remaining algorithms are more robust with respect to recall. Compared to \textsf{BCl}, \textsf{WNP} reduces recall by just -0.2\%, while increasing precision by 26.8\% and F1 by 19.7\%. Yet, \textsf{BLAST} outperforms \textsf{WEP} with respect to all effectiveness measures: recall, precision and F1 raise by 1.3\%, 13.8\% and 11.5\%, respectively. This means that \textsf{BLAST} is able to discard much more non-matching pairs, while retaining a few more matching ones, too. Given that the weight-based pruning algorithms are crafted for applications that promote recall at the cost of slightly lower precision \cite{DBLP:journals/tkde/PapadakisKPN14,papadakis2016scaling}, we select \textsf{BLAST} as the best one in this category, even though it does not achieve the highest F1, on average.


Among the cardinality-based algorithms, we observe that \textsf{RCNP} is a clear winner, outperforming both \textsf{CEP} and \textsf{CNP}. Compared to the former, it
reduces recall by -1.1\%, increasing precision by 44\% and F1 by 34.4\%; compared to the latter, recall drops by -3.5\%, but precision and F1 raise by 37.5\% and 29.3\%, respectively. Given that cardinality-based pruning algorithms are crafted for applications that promote precision at the cost of slightly lower recall \cite{DBLP:journals/tkde/PapadakisKPN14,papadakis2016scaling}, \textsf{RCNP} constitutes the best choice for this category.

\begin{table}[b]
    \centering
    \scriptsize
    \begin{tabular}{|c|c|c|c|c|}
    \hline
    \textbf{ID} & \textbf{Feature set} & \textbf{Recall} & \textbf{Precision} & \textbf{$F1$}          \\ \hline\hline
    72          & \{CF-IBF, RACCB, JS, RS\}  & .8816 & .1932 & .2892  \\ \hline
    74          & \{CF-IBF, RACCB, JS, NRS\} & .8816 & .1932 & .2892         \\ \hline
    75          & \{CF-IBF, RACCB, JS, WJS\} & .8816 & .1932 & .2892        \\ \hline
    78          & \{CF-IBF, RACCB, RS, NRS\} & .8816 & .1932 & .2892         \\ \hline
    79          & \{CF-IBF, RACCB, RS, WJS\} & .8816 & .1932 & .2892         \\ \hline
    82          & \{CF-IBF, RACCB, NRS, WJS\} & .8816 & .1932 & .2892         \\ \hline
    86          & \{CF-IBF, JS, RS, WJS\} & .8816 & .1932 & .2892 \\ \hline
    89          & \{CF-IBF, JS, NRS, WJS\} & .8816 & .1932 & .2892 \\ \hline
    96          & \{CF-IBF, RS, NRS, WJS\} & .8816 & .1932 & .2892 \\ \hline
    190         & \{CF-IBF, RACCB, JS, RS, NRS, WJS\} & .8816 & .1932 & .2892  \\ \hline
    \end{tabular}
    \vspace{0.5em}
    \caption{The 10 feature sets that achieve the highest F1 with \textsf{BLAST}.}
    \label{tab:topblastfeat}
\end{table}

{\color{\hcolor}\textit{Overall, \textsf{RCNP} constitutes the best choice for cardinality-based pruning algorithms, which are crafted for applications that promote precision at the cost of slightly lower recall \cite{DBLP:journals/tkde/PapadakisKPN14,papadakis2016scaling}.
		\textsf{BLAST} is the best among the weight-based pruning algorithms, which are crafted for applications that promote recall at the cost of slightly lower precision \cite{DBLP:journals/tkde/PapadakisKPN14,papadakis2016scaling}.}
}
Note that the F1 of \textsf{BLAST} and \textsf{RCNP} is significantly higher than the original ones in Table \ref{tab:blocking_results}. They are still far from a perferct F1, but (Supervised) Meta-blocking merely produces a new block collection, not the end result of ER. This block collection is then processed by a Matching algorithm, whose goal is to raise F1 close to 1. 



\subsection{Feature selection} \label{sub:feature_selection}

The goal of this experiment is to fine-tune the selected algorithms, namely
\textsf{BLAST} and \textsf{RCNP}, by identifying the feature sets that optimize their performance in terms of both effectiveness and time-efficiency.
To this end, we adopted a brute force approach, trying all the possible combinations of the eight features discussed in Section \ref{sec:weightingSchemes}. Fixing again 
the training set size to 500 random samples, equally split between positive and negative instances, the top-10 feature vectors with respect to 
F1 for \textsf{BLAST} and \textsf{RCNP} are reported in 
Table \ref{tab:topblastfeat} and
Table \ref{tab:top10featuresRCNP}, respectively.

We observe that for each algorithm, all feature sets achieve practically identical performance, on average. 
For \textsf{BLAST}, we obtain recall=$0.882$, precision=$0.193$ and F1=$0.289$, while for \textsf{RCNP}, we obtain recall=$0.850$, precision=$0.248$ and F1=$0.353$. 
Compared to their average performance with the original feature vector proposed in \cite{papadakis2014supervised}, we observe that the average recall and precision raise by $\sim$0.5\%, while the average F1 increases by $~\sim$1.5\%.
This means that the \textit{effectiveness} of both algorithms is quite robust with respect to the underlying feature set. Therefore, we select the best one for each algorithm based on \textit{time efficiency}.

{\color{\hcolor}
	We observe that both algorithms are robust with respect to the top-10 feature sets, as they all achieve practically identical performance, on average.  
	For \textsf{BLAST}, we obtain recall=$0.882$, precision=$0.193$ and F1=$0.289$ when combining $CF$-$IBF$ and $RACCB$ with any two features from $f$=\{$JS$, $RS$, $NRS$, $WJS\}$; even $RACCB$ can be replaced with a third feature from $f$ without any noticeable impact.
	For \textsf{RCNP}, we obtain recall=$0.850$, precision=$0.248$ and F1=$0.353$ when combining $CF$-$IBF$, $RACCB$ and $LCP$ with any pair of features from \{$JS$, $RS$, $NRS$, $WJS$\}.
	In this context, we select the best feature set for each algorithm based on \textit{time efficiency}.
}

In more detail, we compare the top-10 feature sets per algorithm in terms of their running times. This includes
the time required for calculating the features per candidate pair and for retrieving the corresponding classification probability (we exclude the time required for producing the new block collections, because this is a fixed overhead common to all feature sets of the same algorithm).
Due to space limitations, we consider only the two datasets with the most candidate pairs, as reported in Table \ref{tab:datasets}: \texttt{movies} and \texttt{WalmartAmazon}. We repeated every experiment 10 times and took the mean time.

\begin{table}[b]
    \centering
    \scriptsize
    \setlength{\tabcolsep}{4.5pt}
    \begin{tabular}{|c|c|c|c|c|}
    \hline
    \textbf{ID} & \textbf{Feature Set} & \textbf{Recall} & \textbf{Precision} & \textbf{$F1$} \\ \hline\hline
    184         & \{CF-IBF, RACCB, JS, LCP, RS\} & .8489 & .2463 & .3527           \\ \hline
    187         & \{CF-IBF, RACCB, JS, LCP, WJS\}  & .8490 & .2464 & .3526          \\ \hline
    193         & \{CF-IBF, RACCB, LCP, RS, NRS\} & .8490 & .2463 & .3526          \\ \hline
    200         & \{CF-IBF, JS, LCP, RS, NRS\} & .8488 & .2474 & .3526             \\ \hline
    227         & \{CF-IBF, RACCB, JS, LCP, RS, NRS\} & .8493 & .2473 & .3537      \\ \hline
    228         & \{CF-IBF, RACCB, JS, LCP, RS, WJS\} & .8494 & .2473 & .3537      \\ \hline
    231         & \{CF-IBF, RACCB, JS, LCP, NRS, WJS\} & .8496 & .2473 & .3537     \\ \hline
    235         & \{CF-IBF, RACCB, LCP, RS, NRS, WJS\}  & .8496 & .2473 & .3536    \\ \hline
    239         & \{CF-IBF, JS, LCP, RS, NRS, WJS\}  & .8494 & .2473 & .3534       \\ \hline
    \multirow{ 2}{*}{250}         & \{CF-IBF, RACCB, JS, & \multirow{ 2}{*}{.8502} & \multirow{ 2}{*}{.2479} & \multirow{ 2}{*}{.3542} \\ 
    & LCP, RS, NRS, WJS\} & & & \\ 
    \hline
    \end{tabular}
    \vspace{0.5em}
    \caption{The 10 feature sets that achieve the highest F1 when applied to \textsf{RCNP}.}
    \label{tab:top10featuresRCNP}
\end{table}

\begin{figure}[t]
\centering
\includegraphics[width=\linewidth]{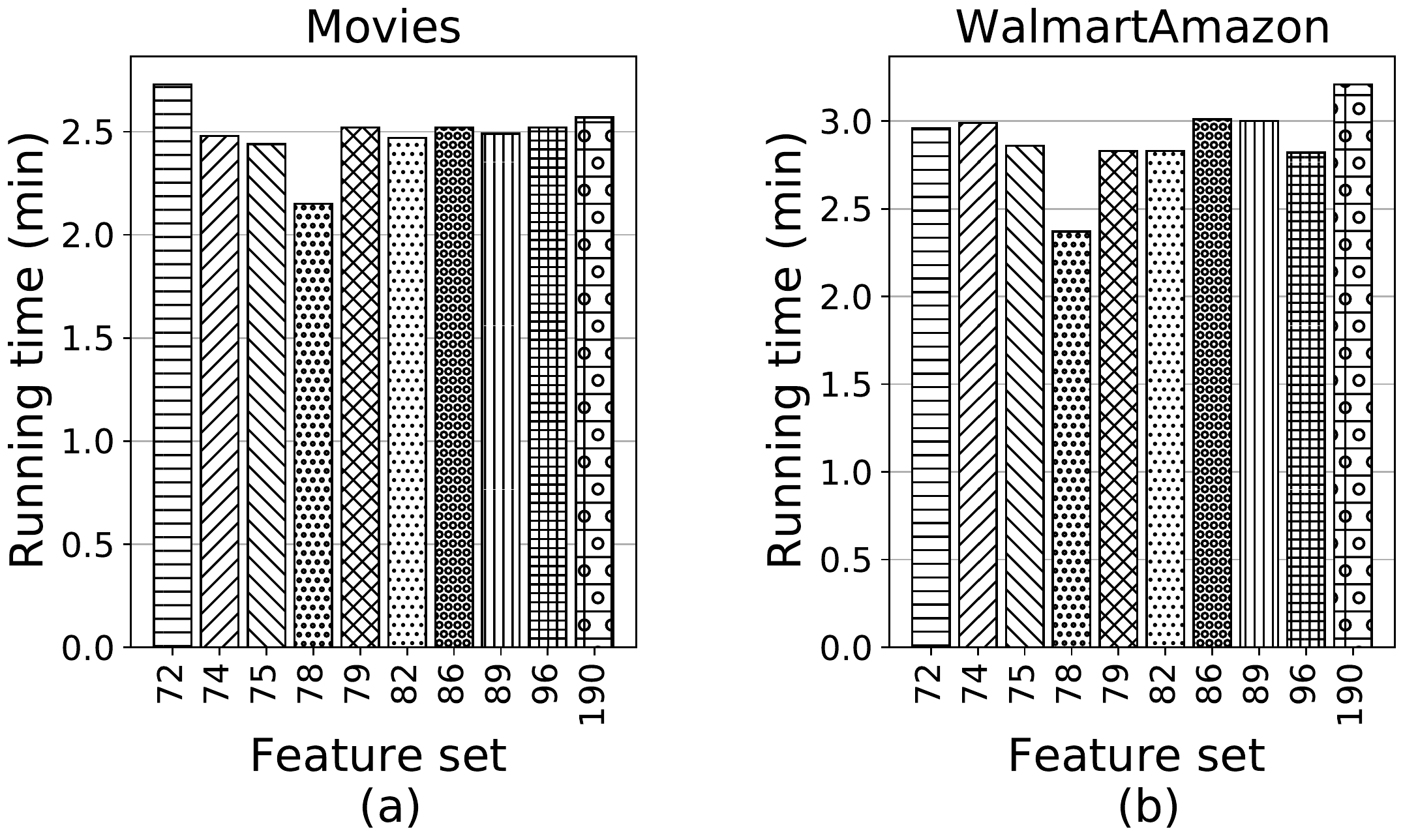}
\vspace{-15pt}
\caption{Running time of top-10 features sets when applied to \textsf{BLAST}.}
\label{fig:exec_time_blast}
\vspace{-10pt}
\end{figure}

\begin{figure}[t]
\centering
\includegraphics[width=\linewidth]{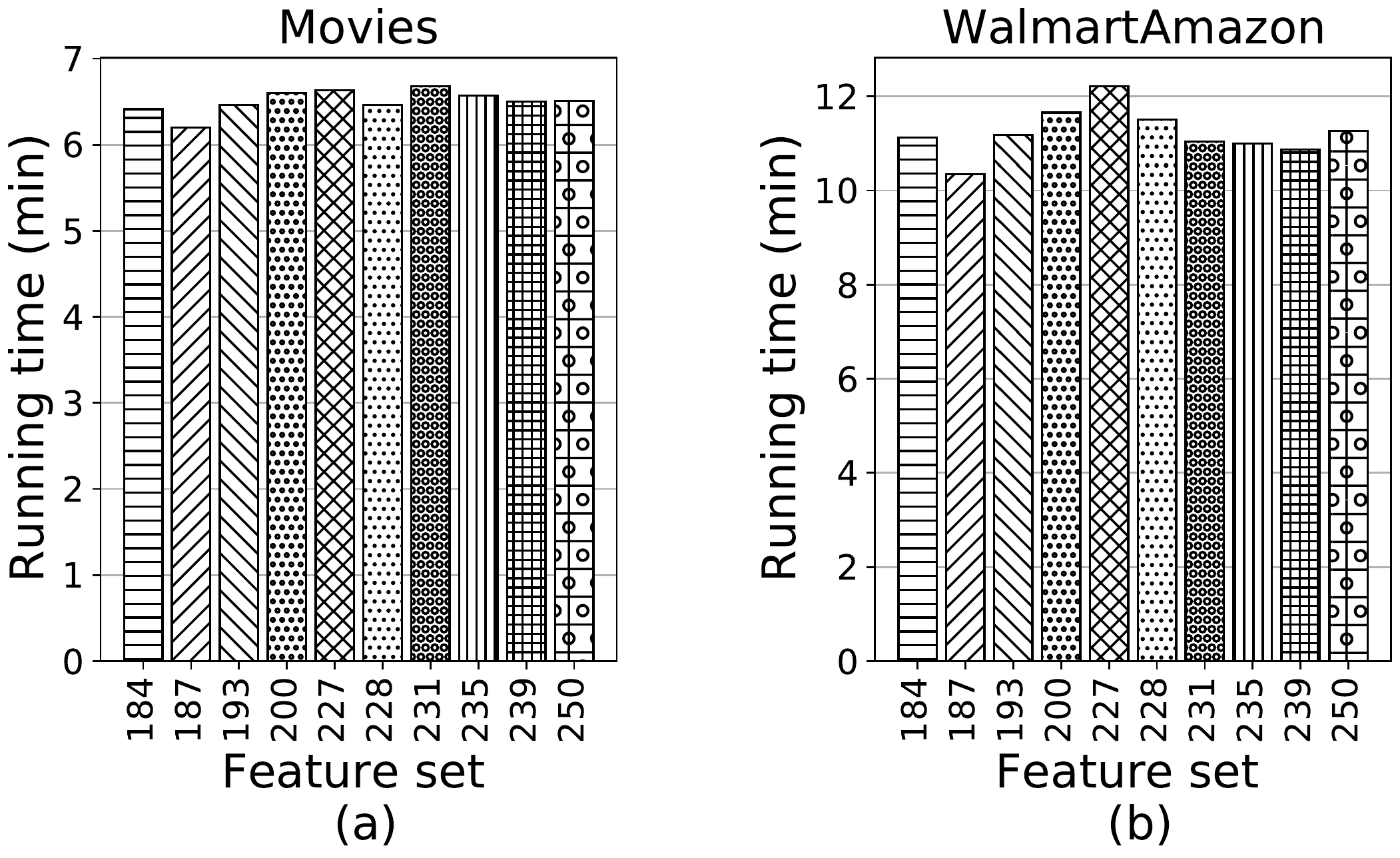}
\vspace{-15pt}
\caption{Running time of top-10 features sets when applied to \textsf{RCNP}.}
\vspace{-10pt}
\label{fig:exec_time_rcnp}
\end{figure}

The resulting running times appear in 
Figures \ref{fig:exec_time_blast} and \ref{fig:exec_time_rcnp} for \textsf{BLAST} and \textsf{RCNP}, respectively.
In the former figure, we observe that the feature set \textbf{78} is consistently the fastest one for \textsf{BLAST}, exhibiting a clear lead. Compared to the second fastest feature sets over \texttt{movies} (75) and \texttt{WalmartAmazon} (96), it reduces the average run-time by 11.9\% and 16.0\%, respectively. For \textsf{RCNP}, the differences are much smaller, yet the same feature set (\textbf{187}) achieves the lowest run-time over both datasets. Compared to the second fastest feature sets over \texttt{movies} (184) and \texttt{WalmartAmazon} (239), it reduces the average run-time by 3.3\% and 4.8\%, respectively.

{\color{\hcolor}
	\textit{Overall, \textsf{BLAST} models each candidate pair as the 4-dimensional feature vector (ID 78 in Table \ref{tab:topblastfeat}a): $\{CF$-$IBF, RACCB, RS, NRS\}$. Compared to the feature set of \cite{papadakis2014supervised}, recall raises by $\sim$0.5\% and F1 by $\sim$1.5\%, while the run-time is reduced to a significant extent ($>$50\% as explained below), due to the absence of the time-consuming $LCP$ feature.
		\textsf{RCNP} represents every candidate pair with the 5-dimensional 
		feature vector (ID 187 in Table \ref{tab:topblastfeat}b): $\{CF$-$IBF, RACCB, JS, LCP, WJS\}$. This reduces recall by $<$0.3\%, but raises precision and F-Measure by 1.2\%, which is in-line with the desiderata of cardinality-based algorithms.
	}
}

The actual features comprising the selected feature sets, 78 and 187, appear in Tables \ref{tab:topblastfeat} and
\ref{tab:top10featuresRCNP}, respectively. In the former, we observe that
\textsf{BLAST} models each candidate pair as the 4-dimensional feature vector:
\begin{equation}
\{CF-IBF, RACCB, RS, NRS\}.
\label{feat:BLAST}
\end{equation}
In the latter table, we observe that \textsf{RCNP} represents every candidate pair with the 5-dimensional 
feature vector:
\begin{equation}
\{CF-IBF, RACCB, JS, LCP, WJS\}.
\label{feat:RCNP}
\end{equation}

It is worth noting that \textsf{BLAST} works better with a smaller feature set than \textsf{RCNP}. In general, all top-10 feature sets selected for the former algorithm are much simpler than those selected for the latter. Most importantly, \textsf{BLAST} avoids the feature $LCP$, whose calculation is quite expensive (it iterates over all blocks containing an entity in order to gather its distinct candidates). 
Therefore, it is no surprise that
\textsf{BLAST} is 2 to 3 times faster than \textsf{RCNP}.




\begin{figure}[b]
\centering
\includegraphics[width=\linewidth]{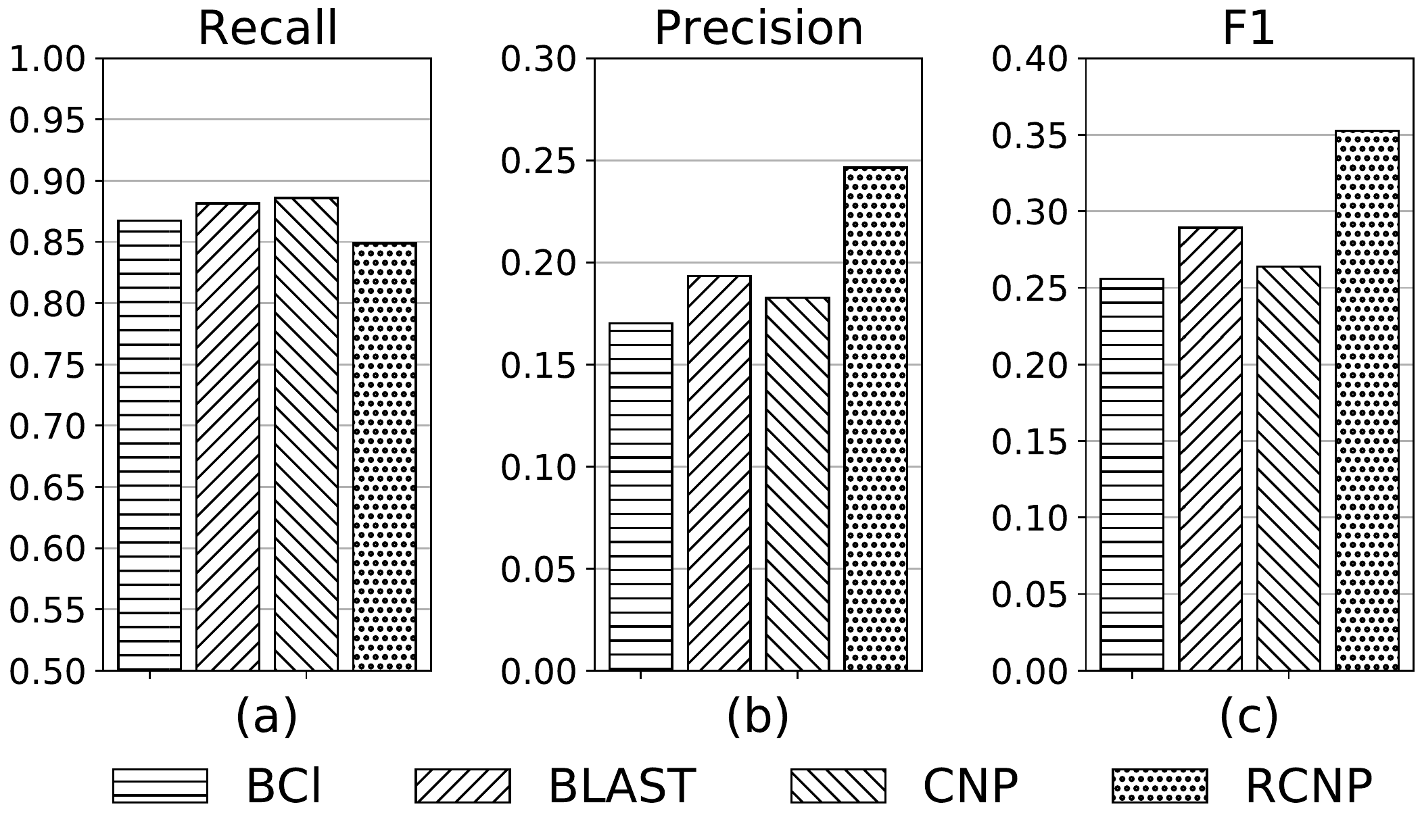}
\vspace{-15pt}
\caption{Comparison of the best algorithms for Supervised (BCl, CNP) and Generalized Supervised Meta-blocking (BLAST, RCNP).}
\label{fig:prec_rec_comp}
\end{figure}

\subsubsection{Comparison with Supervised Meta-blocking \cite{papadakis2014supervised}}
Based on the selected feature sets, we compare the performance of Generalized Supervised Meta-blocking with Supervised Meta-blocking. The former is represented by \textsf{BLAST} and \textsf{RCNP} in combination with the features in Formulas \ref{feat:BLAST} and \ref{feat:RCNP}, respectively, while for the latter,  
we use the feature set proposed in \cite{papadakis2014supervised}, $\{CF-IBF, RACCB, JS, LCP\}$, in combination with \textsf{BCl} and \textsf{CNP}.
All algorithms were trained over a randomly selected set of 500 labelled instances, 250 from each class, and were applied to all datasets in Figure \ref{tab:datasets}. Their average performance with respect to effectiveness is presented in Figure~\ref{fig:prec_rec_comp}.

Among the weight-based algorithms, \textsf{BLAST} achieves higher recall than \textsf{BCl}, by 1.6\% on average. Thus, \textsf{BLAST} is more suitable for recall-intensive applications. Most importantly, it outperforms \textsf{BCl} with respect to the other measures, too: the average precision raises by 13.6\% and the average F1 by 13\%. This indicates that \textsf{BLAST} is much more accurate in the classification of the candidate pairs.

Among the cardinality-based algorithms, \textsf{RCNP} trades slightly lower recall than \textsf{CNP} for significantly higher precision and F1: on average, across all datasets, its recall is lower by -4.1\%, while its precision and F1 are higher by 34.9\% and by 33.6\%, respectively. As a result, \textsf{RCNP} is more suitable for precision-intensive applications. 

\begin{figure}[b]
\centering
\includegraphics[width=\linewidth]{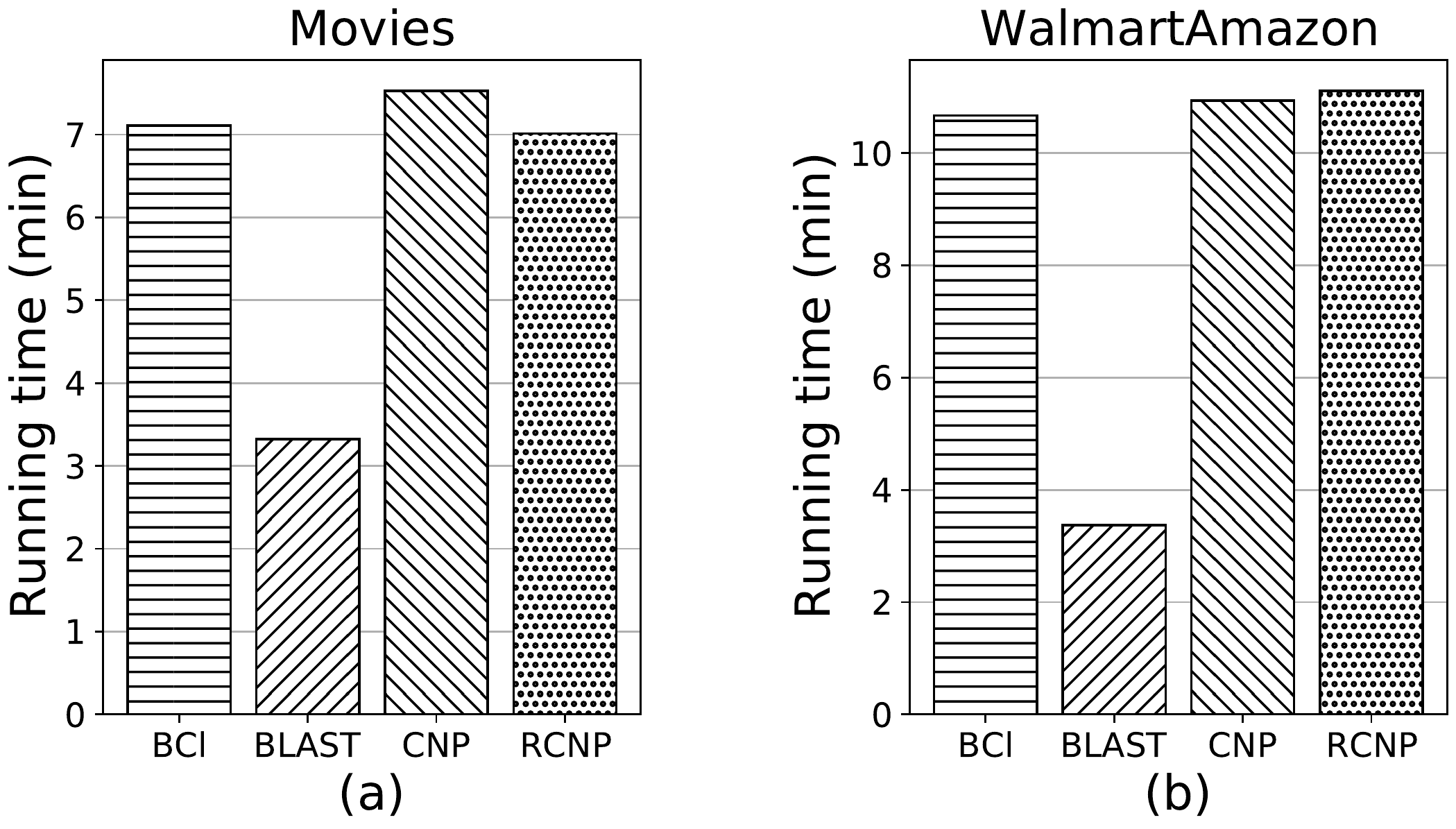}
\vspace{-15pt}
\caption{Comparison of the best algorithms for Supervised (BCl, CNP) and Generalized Supervised Meta-blocking (BLAST, RCNP).}
\label{fig:exec_time_comp}
\end{figure}

Regarding time efficiency, Figure \ref{fig:exec_time_comp} reports the running times of these algorithms 
on the largest datasets, i.e., \texttt{Movies} and \texttt{WalmartAmazon}. We observe that \textsf{BCl}, \textsf{CNP} and \textsf{RCNP} exhibit similar $RT$ in both cases, since they all employ more complex feature sets that include the time-consuming feature $LCP$.
\textsf{BLAST} is substantially faster than these algorithms, reducing $RT$ by more than 50\%.
In particular, comparing it with its weight-based competitor, we observe that \textsf{BLAST} 
is faster than \textsf{BCl} by 2.1 times over \texttt{Movies} and by 3.2 times over \texttt{WalmartAmazon}.

We can conclude, therefore, that Generalized Supervised Meta-blocking conveys significant improvements with respect to Supervised Meta-blocking.


\subsection{The effect of training set size}
\label{sec:trainingSet}

\begin{figure}[t]
\centering
\includegraphics[width=\linewidth]{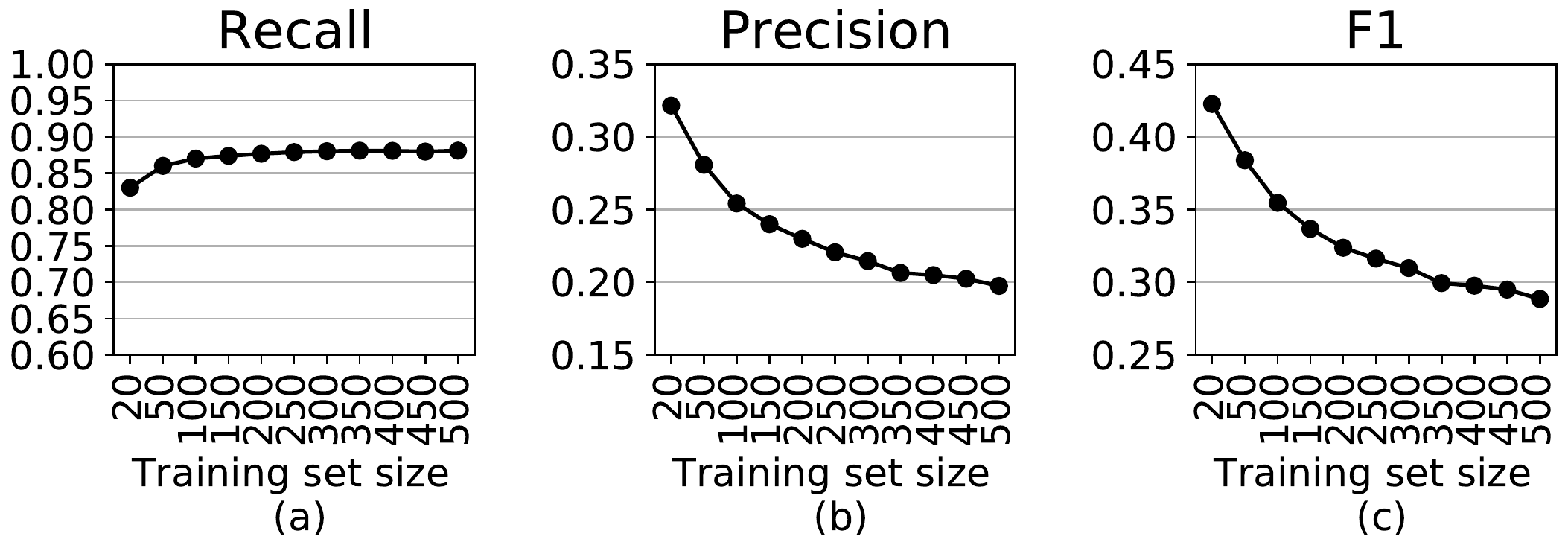}
\vspace{-15pt}
\caption{The effect of the training set size on \textsf{BLAST}.}
\vspace{-15pt}
\label{fig:blast_train_size}
\end{figure}

\begin{figure}[t]
\centering
\includegraphics[width=\linewidth]{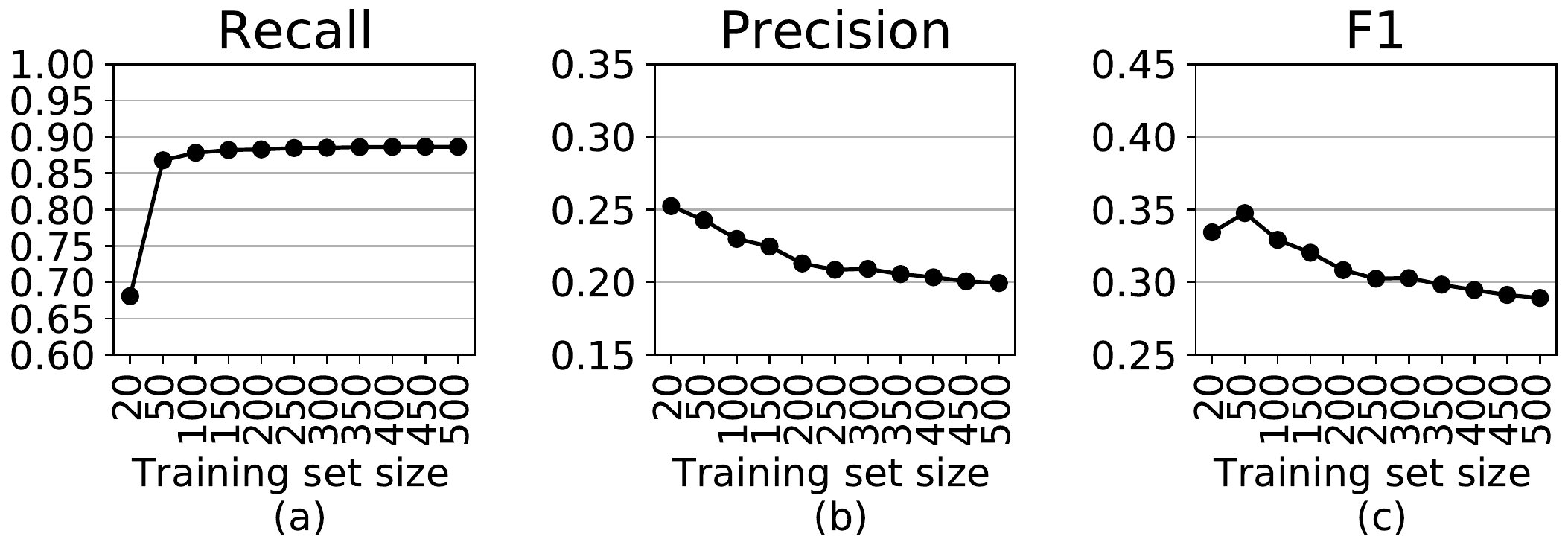}
\vspace{-15pt}
\caption{The effect of the training set size on \textsf{RCNP}.}
\vspace{-15pt}
\label{fig:rcnp_train_size}
\end{figure}

To examine whether active learning is necessary for \textsf{BLAST} and \textsf{RCNP}, we perform an 
experiment that explores how their performance 
changes when varying the training set size.
We used the features sets specified in Formulas \ref{feat:BLAST} and \ref{feat:RCNP} and varied the number of labelled instances {\color{\hcolor}from 20, then from 50 to 500 with a step of 50.}\footnote{Note that we tried to use BLOSS \cite{DBLP:journals/is/BiancoGD18} as a baseline method, but we couldn't reproduce its performance, since our implementation of the algorithm exclusively selected non-matching candidate pairs -- instead of a balanced training set. We contacted the authors, but they were not able to provide us with their own implementation. Nevertheless, our experimental results demonstrate that active learning is not necessary for our approaches, given that they achieve high performance with just 50 labelled instances.}
Figures \ref{fig:blast_train_size} and \ref{fig:rcnp_train_size} report the results in terms of recall, precision and F1, on average across all datasets, for \textsf{BLAST} and \textsf{RCNP}, respectively. 

Notice that both algorithms exhibit the same behavior: as the training set size increases, recall gets higher at the expense of lower precision and F1. However, the increase in recall is much lower than the decrease in the other two measures. More specifically, comparing the largest training set size with the smallest one, the average recall of \textsf{BLAST} raises by 2.4\%, while its average precision drops by 29.7\% and its average F1 by 24.8\%. Similar patterns apply to \textsf{RCNP}: recall raises by 2.1\%, but precision and F1 drop by 17.8\%	and 16.8\%, respectively, when increasing the labelled instances from 50 to 500. 
{\color{\hcolor}This might seem counter-intuitive, 
larger training sets are expected to improve performance by increasing both recall and precision.
In our cases, only recall raises with higher training sets, unlike precision.
This behavior should be attributed to the distribution of the matching probabilities that are produced by the trained probabilistic classifier.
In more detail, with larger training sets, the matching probabilities of duplicate pairs are pushed up, thus raising recall.
The same applies to the probabilities of non-matching pairs, though, thus decreasing precision. 

This is illustrated in Figure \ref{fig:abtBuyProbabilties}, which depicts the density of matching probabilities of candidate pairs over the AbtBuy dataset for various training set sizes, when using the {\color{red}Logistic Regression} as the classification algorithm. 
The duplicate pairs are shown in red and the non-matching ones in blue. The upper line corresponds to the maximum pruning threshold and the lower line to the average one, across all nodes/entities. We observe that for the smallest training set, the matching probabilities of both types of candidate pairs fluctuate in $[0.5, 0.95]$, with the density of the matching and the non-matching pairs being higher in the upper and the lower part, respectively.
As we move to larger training sets, the matching pairs are pushed up, fluctuating in $[0.75, 0.95]$.
This indicates the higher discriminatory power of the probabilistic classifier, which increases the number of true positives, raising recall. However, the probabilities of the non-matching pair continue to fluctuate in $[0.5, 0.85]$, but are now concentrated around {\color{red}0.7}, while the most pruning thresholds are confined in $[0.63, 0.65]$. As a result, the number of false positive increases, too, dropping precision to lower levels.
Similar patterns appear in the rest of the datasets in combination with other classification algorithms, such as {\color{red}SVC}. 

It is worth noting at this point that the same behavior as BLAST is exhibited by BCl, the original binary classifier presented in \cite{papadakis2014supervised}, which simply retains all candidate pairs with a matching probability above 0.5. As shown in Figure \ref{fig:bclRecallPrecision}, both algorithms increase their recall and decrease their precision as more labelled instances as used during their training.

\begin{figure}[H]
	\centering
	\includegraphics[width=\linewidth]{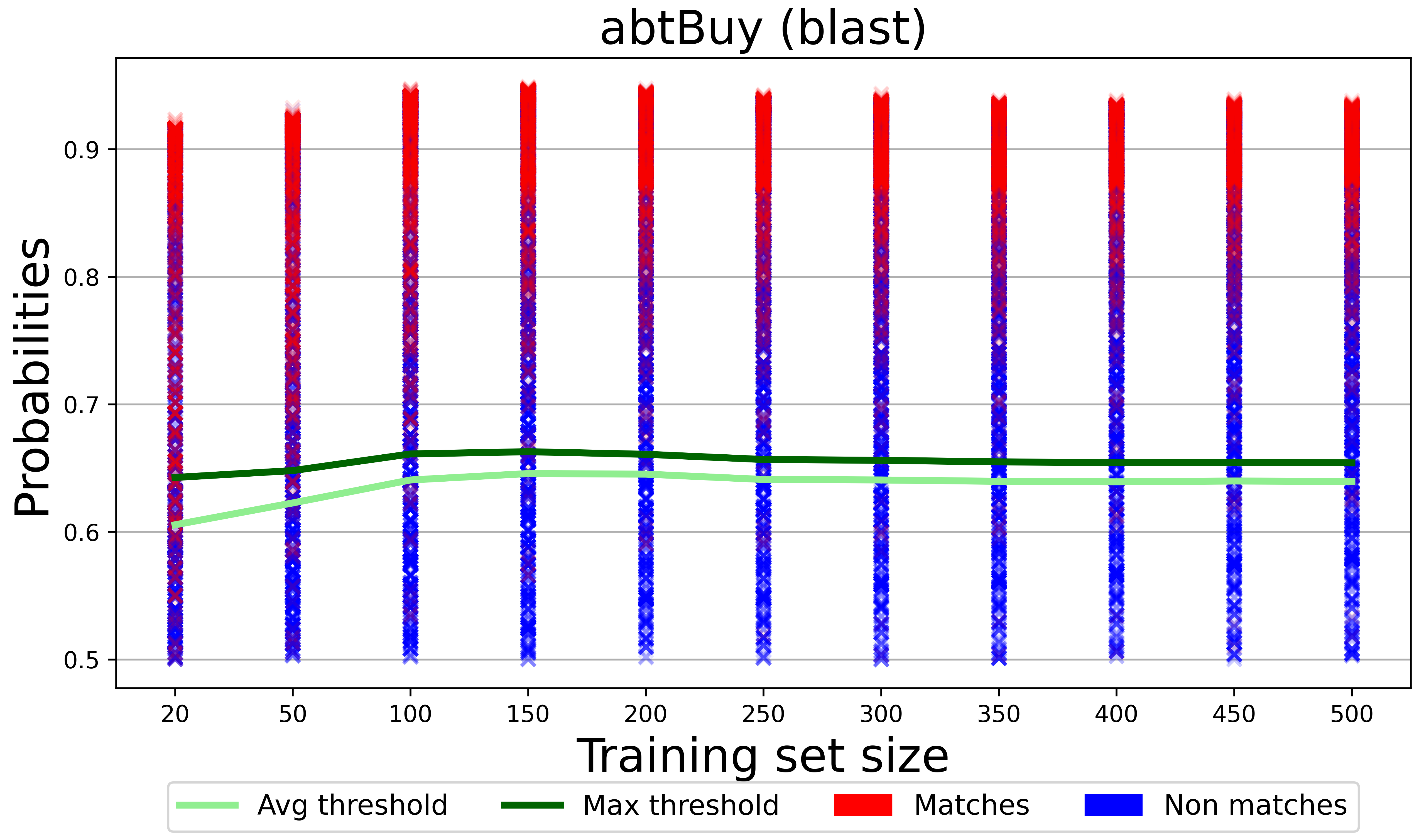}
	\caption{The matching probabilities for both types of candidate pairs, the duplicate and the non-matching ones, as the size of the training set increases.}
	\label{fig:abtBuyProbabilties}
\end{figure}

\begin{figure}[H]
	\centering
	\includegraphics[width=\linewidth]{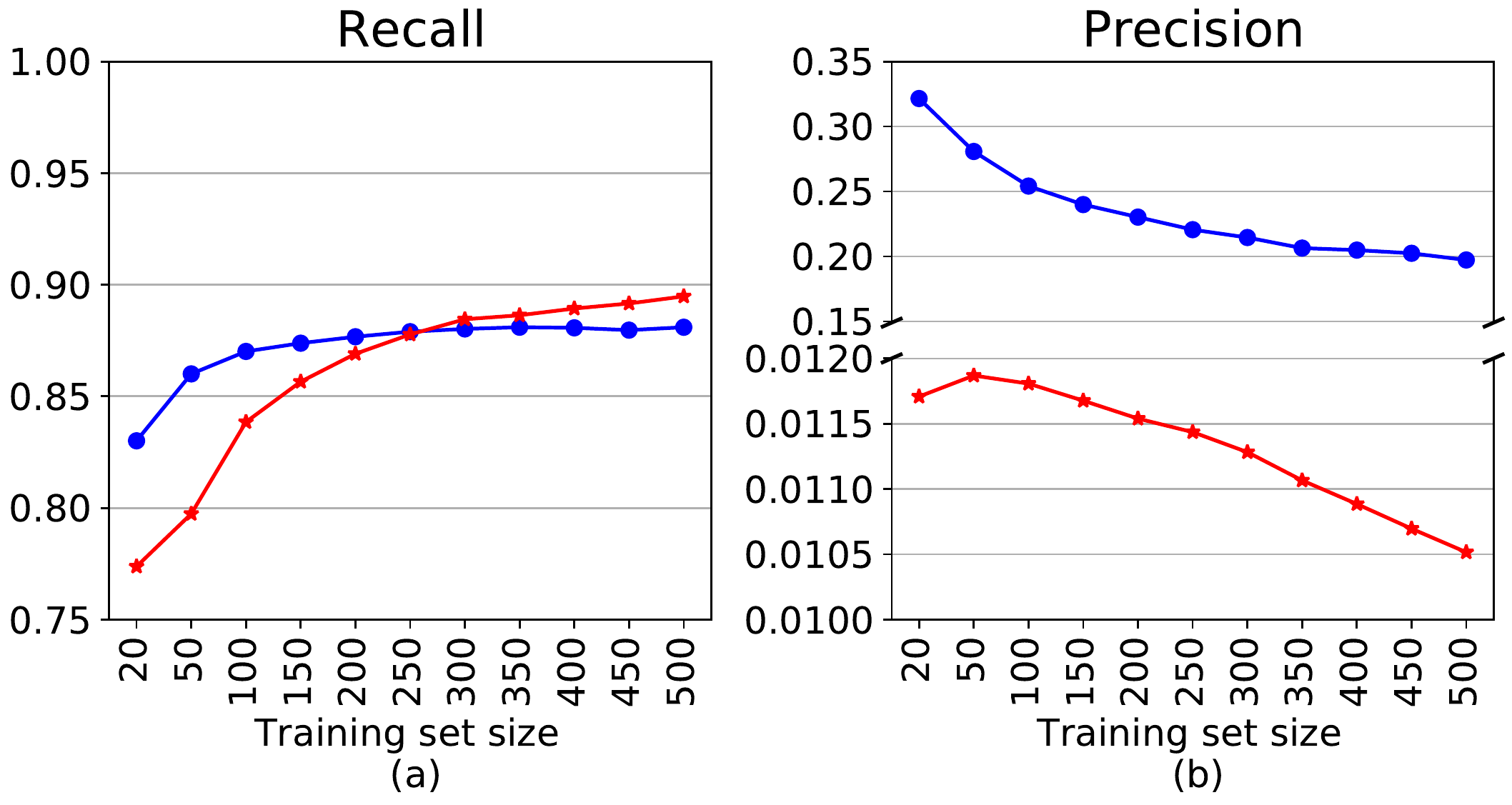}
	\caption{Recall and precision of \textsf{BCl} and \textsf{BLAST} as the size of the training set increases.}
	\label{fig:bclRecallPrecision}
\end{figure}
	
}

Given that the initial levels of recall are quite satisfactory for both algorithms ($\gg$0.85, on average, across all datasets), we can conclude that \textit{the optimal training set involves just 50 labelled instances, equally split among positive and negative ones}. This renders active learning techniques unnecessary, given that their cost is not compensated, when such a small training set is required.

\begin{table*}[h]
    \footnotesize
    \centering
    \begin{tabular}{lccccccccc}
    \cline{2-10}
    \multicolumn{1}{c|}{}                      & \multicolumn{1}{c|}{\textbf{AbtBuy}} & \multicolumn{1}{c|}{\textbf{DblpAcm}} & \multicolumn{1}{c|}{\textbf{ScholarDblp}} & \multicolumn{1}{c|}{\textbf{AmazonGP}} & \multicolumn{1}{c|}{\textbf{ImdbTmdb}} & \multicolumn{1}{c|}{\textbf{ImdbTvdb}} & \multicolumn{1}{c|}{\textbf{TmdbTvdb}} & \multicolumn{1}{c|}{\textbf{Movies}} & \multicolumn{1}{c|}{\textbf{WMAmazon}} \\ 
    \hline\hline
    \multicolumn{1}{|l|}{$Re$}      & \multicolumn{1}{c|}{0.8345}          & \multicolumn{1}{c|}{0.9511}           & \multicolumn{1}{c|}{0.9638}               & \multicolumn{1}{c|}{0.7001}                        & \multicolumn{1}{c|}{0.8223}            & \multicolumn{1}{c|}{0.7483}            & \multicolumn{1}{c|}{0.8466}            & \multicolumn{1}{c|}{0.9151}          & \multicolumn{1}{c|}{0.9587}                 \\
    \multicolumn{1}{|l|}{$Pr$}   & \multicolumn{1}{c|}{0.2037}          & \multicolumn{1}{c|}{0.6509}           & \multicolumn{1}{c|}{0.3418}               & \multicolumn{1}{c|}{0.1441}                        & \multicolumn{1}{c|}{0.5756}            & \multicolumn{1}{c|}{0.2304}            & \multicolumn{1}{c|}{0.2477}            & \multicolumn{1}{c|}{0.1300}          & \multicolumn{1}{c|}{0.0025}                 \\
    \multicolumn{1}{|l|}{$F1$}          & \multicolumn{1}{c|}{0.3265}          & \multicolumn{1}{c|}{0.7690}           & \multicolumn{1}{c|}{0.4988}               & \multicolumn{1}{c|}{0.2385}                        & \multicolumn{1}{c|}{0.6726}            & \multicolumn{1}{c|}{0.3456}            & \multicolumn{1}{c|}{0.3770}            & \multicolumn{1}{c|}{0.2221}          & \multicolumn{1}{c|}{0.0050}                 \\
    \multicolumn{1}{|l|}{{$RT$ (sec)}} & \multicolumn{1}{c|}{6.58}            & \multicolumn{1}{c|}{5.62}             & \multicolumn{1}{c|}{11.90}                & \multicolumn{1}{c|}{6.83}                          & \multicolumn{1}{c|}{6.46}              & \multicolumn{1}{c|}{6.36}              & \multicolumn{1}{c|}{7.51}              & \multicolumn{1}{c|}{96.01}           & \multicolumn{1}{c|}{107.82}                 \\ \hline
    \multicolumn{10}{c}{\textbf{(a) BLAST in combination with 50 balanced labelled instances and $\{CF$-$IBF, RACCB, RS, NRS\}$ }}                                                                                                                                                                                                                                                                                                                                                                    \\ \hline
    \multicolumn{1}{|l|}{$Re$}      & \multicolumn{1}{c|}{\color{\hcolor}0.8345}          & \multicolumn{1}{c|}{\color{\hcolor}0.9521}           & \multicolumn{1}{c|}{\color{\hcolor}0.9588}               & \multicolumn{1}{c|}{\color{\hcolor}0.6265}                        & \multicolumn{1}{c|}{\color{\hcolor}0.7889}            & \multicolumn{1}{c|}{\color{\hcolor}0.6966}            & \multicolumn{1}{c|}{\color{\hcolor}0.6972}            & \multicolumn{1}{c|}{\color{\hcolor}0.9039}          & \multicolumn{1}{c|}{\color{\hcolor}0.9500}                 \\
    \multicolumn{1}{|l|}{$Pr$}   & \multicolumn{1}{c|}{\color{\hcolor}0.1821}          & \multicolumn{1}{c|}{\color{\hcolor}0.5971}           & \multicolumn{1}{c|}{\color{\hcolor}0.3595}               & \multicolumn{1}{c|}{\color{\hcolor}0.1607}                        & \multicolumn{1}{c|}{\color{\hcolor}0.6445}            & \multicolumn{1}{c|}{\color{\hcolor}0.2616}            & \multicolumn{1}{c|}{\color{\hcolor}0.3737}            & \multicolumn{1}{c|}{\color{\hcolor}0.0972}          & \multicolumn{1}{c|}{\color{\hcolor}0.0020}               \\
    \multicolumn{1}{|l|}{$F1$}          & \multicolumn{1}{c|}{\color{\hcolor}0.2981}          & \multicolumn{1}{c|}{\color{\hcolor}0.7303}           & \multicolumn{1}{c|}{\color{\hcolor}0.5195}               & \multicolumn{1}{c|}{\color{\hcolor}0.2572}                        & \multicolumn{1}{c|}{\color{\hcolor}0.7086}            & \multicolumn{1}{c|}{\color{\hcolor}0.3785}            & \multicolumn{1}{c|}{\color{\hcolor}0.4613}            & \multicolumn{1}{c|}{\color{\hcolor}0.1735}          & \multicolumn{1}{c|}{\color{\hcolor}0.0041}                 \\
    \multicolumn{1}{|l|}{{$RT$ (sec)}} & \multicolumn{1}{c|}{\color{\hcolor}5.40}            & \multicolumn{1}{c|}{\color{\hcolor}5.66}             & \multicolumn{1}{c|}{\color{\hcolor}10.51}                & \multicolumn{1}{c|}{\color{\hcolor}6.02}                          & \multicolumn{1}{c|}{\color{\hcolor}5.79}              & \multicolumn{1}{c|}{\color{\hcolor}5.49}              & \multicolumn{1}{c|}{\color{\hcolor}6.69}              & \multicolumn{1}{c|}{\color{\hcolor}82.71}           & \multicolumn{1}{c|}{\color{\hcolor}107.51}                 \\ \hline
    \multicolumn{10}{c}{\color{\hcolor}\textbf{(b) BCl$_1$ in combination with 50 balanced labelled instances and $\{CF$-$IBF, RACCB, RS, NRS\}$ }}                                                                                                                                                                                                                                                                                                                                                                                       \\ \hline
    \multicolumn{1}{|l|}{$Re$}      & \multicolumn{1}{c|}{0.8183}          & \multicolumn{1}{c|}{0.9513}           & \multicolumn{1}{c|}{0.9303}               & \multicolumn{1}{c|}{0.7316}                        & \multicolumn{1}{c|}{0.7872}            & \multicolumn{1}{c|}{0.7074}            & \multicolumn{1}{c|}{0.8172}            & \multicolumn{1}{c|}{0.9100}          & \multicolumn{1}{c|}{0.5757}                 \\
    \multicolumn{1}{|l|}{$Pr$}   & \multicolumn{1}{c|}{0.2039}          & \multicolumn{1}{c|}{0.6130}           & \multicolumn{1}{c|}{0.3921}               & \multicolumn{1}{c|}{0.1131}                        & \multicolumn{1}{c|}{0.5969}            & \multicolumn{1}{c|}{0.2323}            & \multicolumn{1}{c|}{0.2312}            & \multicolumn{1}{c|}{0.0239}          & \multicolumn{1}{c|}{0.0001}               \\
    \multicolumn{1}{|l|}{$F1$}          & \multicolumn{1}{c|}{0.3261}          & \multicolumn{1}{c|}{0.7425}           & \multicolumn{1}{c|}{0.5401}               & \multicolumn{1}{c|}{0.1908}                        & \multicolumn{1}{c|}{0.6604}            & \multicolumn{1}{c|}{0.3395}            & \multicolumn{1}{c|}{0.2991}            & \multicolumn{1}{c|}{0.0465}          & \multicolumn{1}{c|}{0.0001}                 \\
    \multicolumn{1}{|l|}{$RT$ (sec)} & \multicolumn{1}{c|}{15.07}           & \multicolumn{1}{c|}{9.37}             & \multicolumn{1}{c|}{27.73}                & \multicolumn{1}{c|}{13.22}                         & \multicolumn{1}{c|}{11.04}             & \multicolumn{1}{c|}{9.68}              & \multicolumn{1}{c|}{10.86}             & \multicolumn{1}{c|}{1,328.81}         & \multicolumn{1}{c|}{276.19}                 \\ \hline
    \multicolumn{10}{c}{\textbf{(c) BCl$_2$ in combination with the training set and the features proposed in \cite{papadakis2014supervised}, i.e., $\{CF$-$IBF, RACCB, JS, LCP\}$}}                                                                                                                                                                                                                                                                                                                                                                                              
    \end{tabular}
    \caption{The performance of the main weight-based algorithms across all datasets. 
    $RT$ is the mean run-time over 10 repetitions.}
    \vspace{-20pt}
    \label{tab:full_comp_w_based}
\end{table*}
\subsubsection{Comparison with Supervised Meta-blocking \cite{papadakis2014supervised}}

Based on the above experimental results, we now compare the final algorithms of Generalized Supervised Meta-blocking with their baseline counterparts from \cite{papadakis2014supervised}.

Regarding the weight-based algorithms, Table \ref{tab:full_comp_w_based} reports a full comparison between \textsf{BLAST} and \textsf{BCl}. The former uses the features in Formula \ref{feat:BLAST} along with 50 labelled instances, while the latter combines the best feature set from \cite{papadakis2014supervised}, $\{CF-IBF, RACCB, JS, LCP\}$, with two different training sets: \textsf{BCl}$_1$ uses the same 50 labelled instances that are used by \textsf{BLAST}, while \textsf{BCl}$_2$ uses the training set specified in \cite{papadakis2014supervised} (i.e., a random sample involving 5\% of the positive instances in the ground-truth along with an equal number of randomly selected negative instances).

{\color{\hcolor}
	We observe that on average, \textsf{BLAST} outperforms \textsf{BCl}$_2$ with respect to all effectiveness measures, increasing the average recall, precision and F1 by 7.1\%, 5.0\% and 9.9\%, respectively. Compared to \textsf{BCl}$_1$, \textsf{BLAST} increases the average recall by 3.95\%, at the cost of slightly lower precision and F1 (5.9\% and 2.2\%, respectively). Note \textsf{BCl}$_1$ outperforms \textsf{BCl}$_2$ in all respects,  
	demonstrating the effectiveness of the new 
	feature set. Note also that \textsf{BLAST}'s recall
	is the highest one in all datasets, except \texttt{AmazonGP}, dropping below 0.8 in two datasets (this happens in four datasets for \textsf{BCl}$_1$ and \textsf{BCl}$_2$). This should be attributed to duplicate pairs that share just one block in the original block collection, due to missing or erroneous values, as explained in detail in ~\cite{EXTENDEDVERSION}.
	
	In terms of run-time, \textsf{BLAST} is slower than \textsf{BCl}$_1$ by 8.2\%, on average, because it
	iterates once more over all candidate pairs.
	Compared to \textsf{BCl}$_2$, \textsf{BLAST} is 6.7 times faster, on average across all datasets, because of $LCP$ and of the large training sets, which learn complex binary classifiers with a time consuming~processing. 
} 

{\color{\hcolor}
	Similar patterns are observed in the case of cardinality-based algorithms in Table \ref{tab:full_comp_c_based}.
	\textsf{RCNP} outperforms both baseline methods for all effectiveness measures. Compared to \textsf{CNP}$_1$ (\textsf{CNP}$_2$), \textsf{RCNP} raises the average recall by 5.3\% (9.2\%), the average precision by 0.6\% (16.4\%) and the average F1 by 7.7\% (18.3\%).
	Most importantly, it achieves the highest precision in all datasets, except for \texttt{AbtBuy} and \texttt{ImdbTmdb}. The same applies to F1, too, indicating that recall is not excessively sacrificed in favor of precision. 
	Instead, \textsf{RCNP} is typically more accurate in classifying the positive candidate pairs, since it yields the maximum recall in half the datasets. 
	In terms of run-time, \textsf{RCNP} is slower than \textsf{CNP}$_1$ by 2.5\%, on average, as the latter employs less features, learning simpler and faster classification models than the former when using the same labelled instances. \textsf{CNP}$_2$ employs a much larger training set, yielding more complicated and time-consuming classifiers than \textsf{RCNP}, which is 6.3 times faster,
	on average, across all~datasets.
}

Overall, Generalized Supervised Meta-blocking outperforms Supervised Meta-blocking to a significant extent.

{\color{\hcolor}
	\subsubsection{Considerations regarding the obtained recall.}
	Blocking recall typically sets the upper bound on matching recall, i.e., on the overall recall of ER. 
	The problem of low recall appears in cases where $PC<0.9$ in Table \ref{tab:full_comp_w_based}, which reports analytically the performance over all datasets for the weight-based algorithms that emphasize recall. 
	The rest of the tables report the average recall across all datasets, which is very close to 0.9 for BLAST in most cases, while Table \ref{tab:full_comp_c_based} examines cardinality-based algorithms, which emphasize precision (e.g., progressive ER applications that by default operate with lower recall).
	
	Studying Table \ref{tab:full_comp_w_based}, we observe that the recall of BLAST is lower than 0.9 in five datasets: AbtBuy, AmazonGP, ImdbTmdb, ImdbTvdb and TmdbTvdb. Comparing these datasets with the remaining four, where recall is well above 0.9, we observe that they involve high levels of noise, which causes a large part of the duplicate entities to share just one block. Inevitably, these matching pairs lack significant co-occurrence patterns and receive low probabilities by all weighting schemes used by our approach. As a result, most of them are pruned at the cost of lower recall. In other words, BLAST exhibits low recall in block collections with very low levels of redundancy (due to noisy and missing values), where the matching pairs have just one block in common. Note, though, that Generalized Supervised Meta-blocking outperforms the original Supervised Meta-blocking approach, whose $PC$ is lower by $\sim$5\% in these cases; this means that more duplicates pairs with a single common block are retained by our approach, while increasing precision, too. This should be attributed to the new features we have proposed in this work, which normalize the co-occurrence patterns by considering all blocks containing every entity. In block collections with high levels of redundancy, where the vast majority of duplicates co-occurs in multiple blocks, the recall remains high even after Supervised Meta-blocking.
	
	This is verified in Figures \ref{fig:highRecall} and \ref{fig:lowRecall}, which plot the portion of matching pairs (on the vertical axis) with respect to the number of common blocks (on the horizontal axis). In every dataset, the bar that corresponds to $x=0$ indicates the portion of duplicates that are missed by the original block collection. The bar the corresponds to $x=1$ indicates the portion of duplicates that are missed by (Generalized) Supervised Meta-blocking. We observe that for all datasets in Figure \ref{fig:highRecall} both bars are below 5\%, while for all datasets in Figure \ref{fig:lowRecall}, more than 10\% corresponds to $x=1$. As a result, $PC>0.9$ for the former datasets and vice versa for the latter ones.

	\begin{figure}[t]
		\centering
		\includegraphics[width=\linewidth]{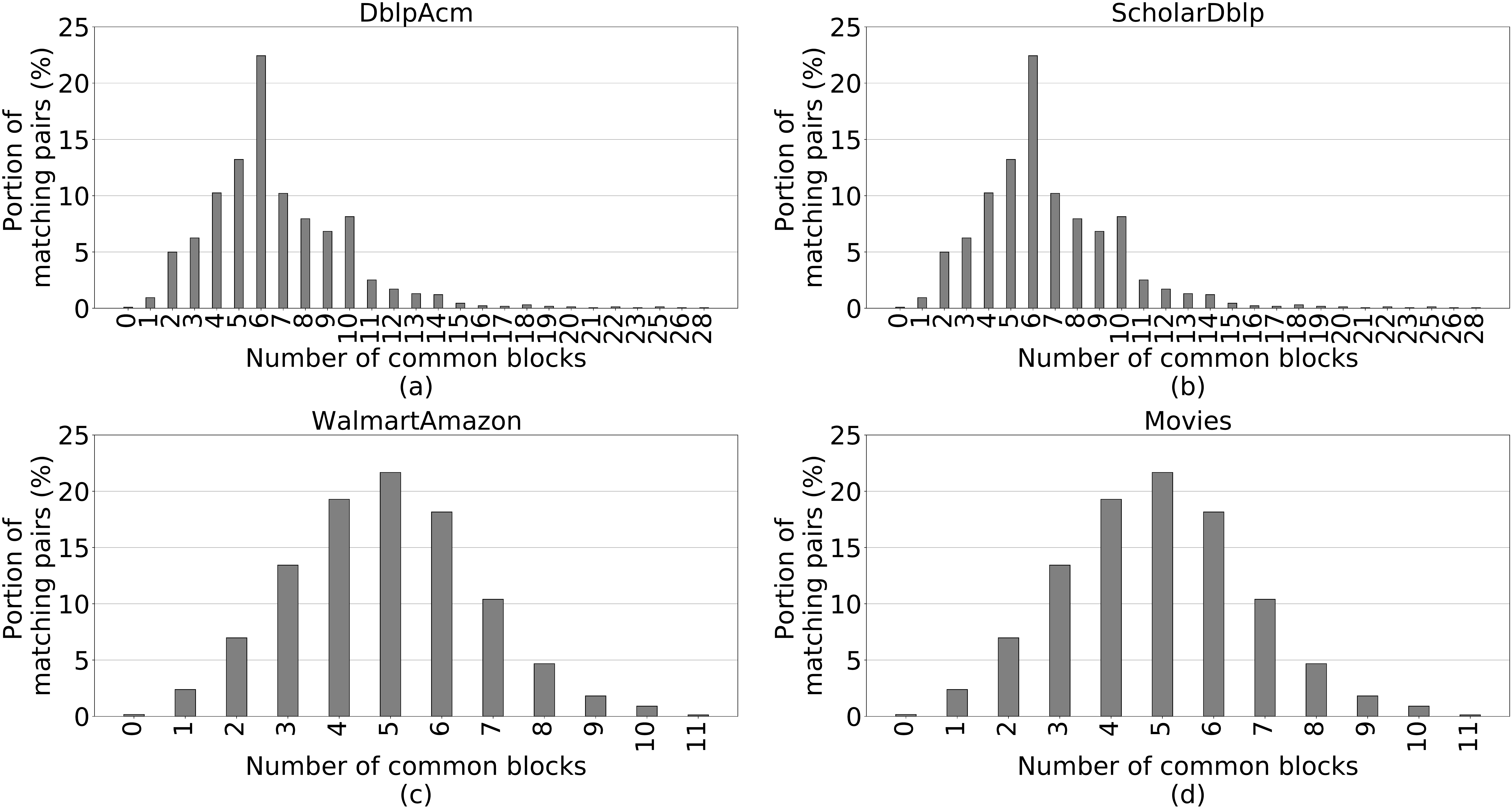}
		\caption{The distribution of common blocks (horizontal axis) per portion of duplicate pairs (vertical axis) in datasets with $PC>0.9$ for BLAST.}
		\label{fig:highRecall}
	\end{figure}
	
	\begin{figure}[t]
		\centering
		\includegraphics[width=\linewidth]{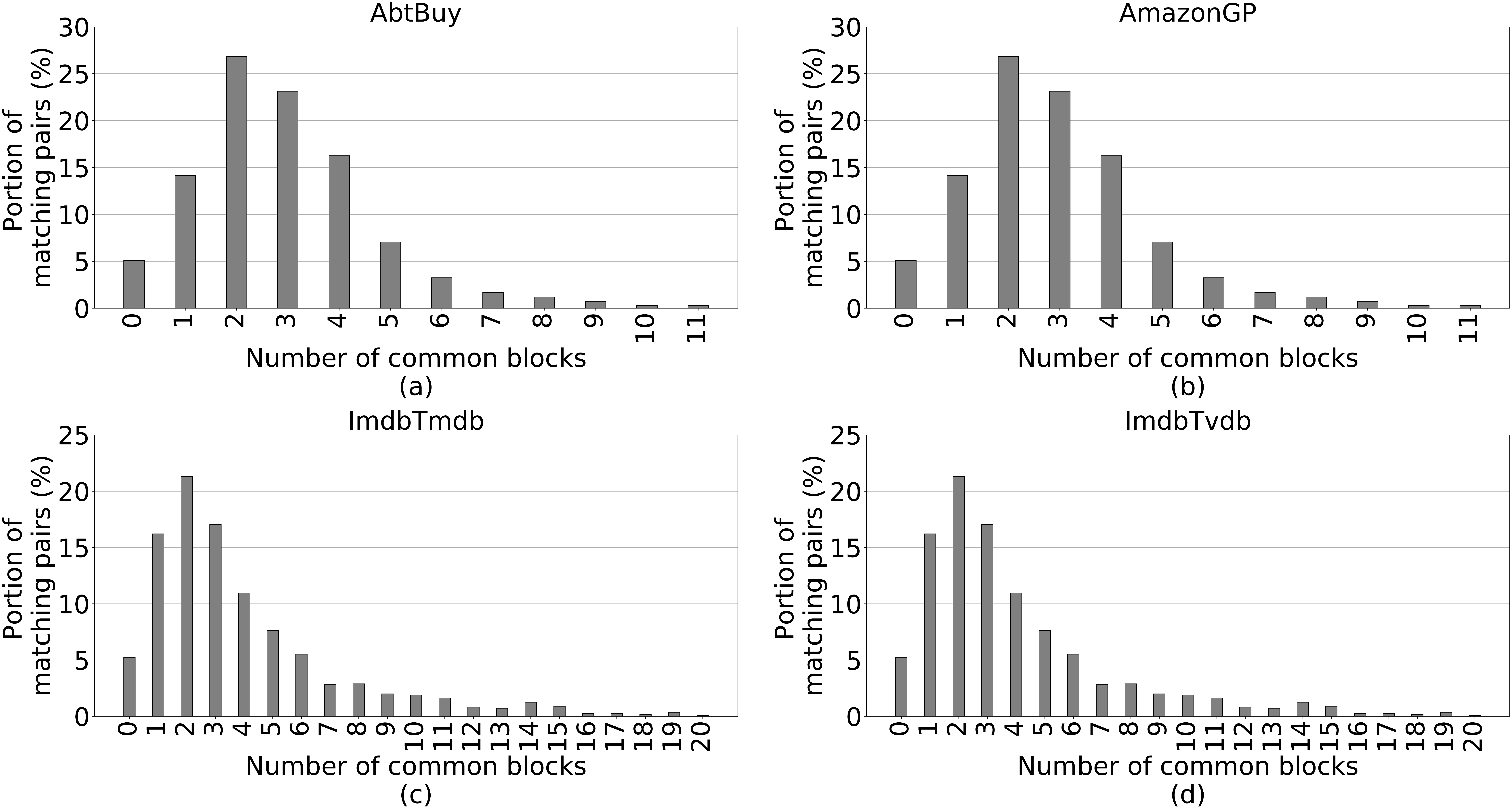}
		\includegraphics[width=\linewidth]{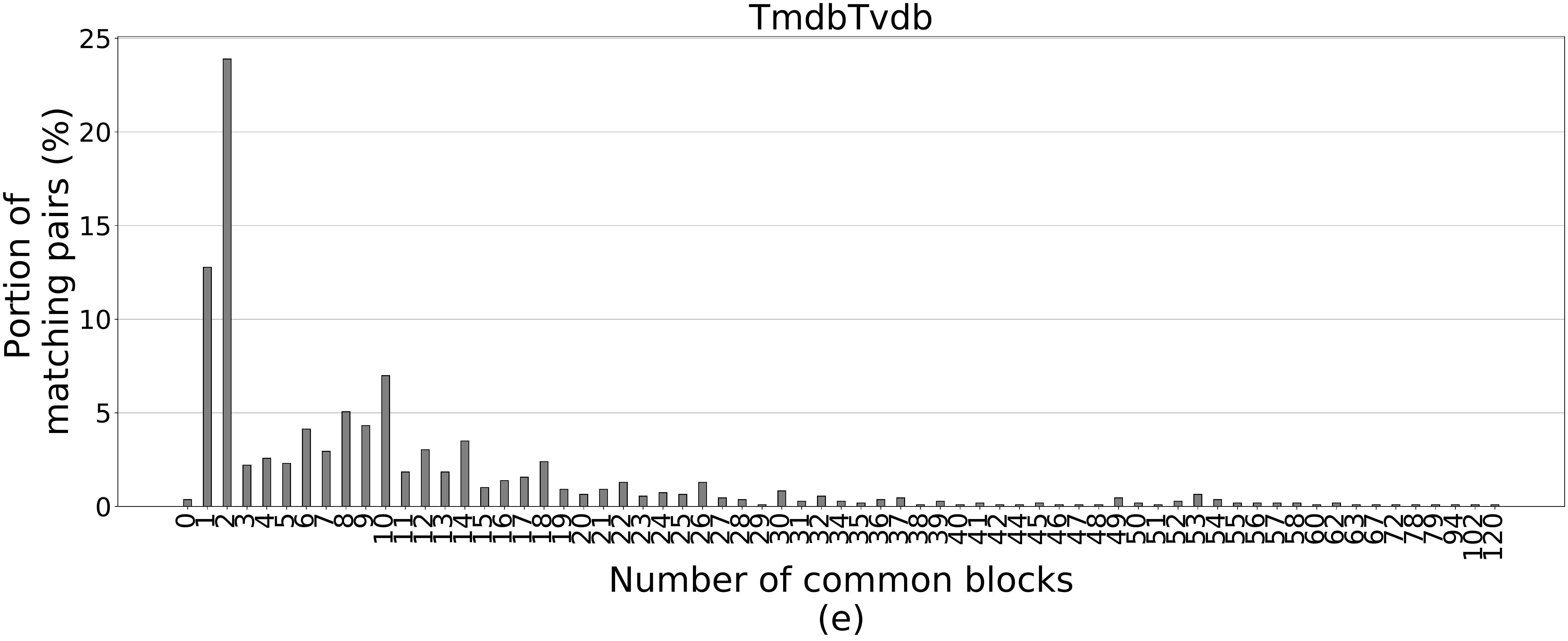}
		\caption{The distribution of common blocks (horizontal axis) per portion of duplicate pairs (vertical axis) in datasets with $PC<0.9$ for BLAST.}
		\label{fig:lowRecall}
	\end{figure}
}

\vspace{-10pt}
\subsection{Scalability Analysis}

We assess the scalability of our approaches as the number of candidate pairs $|C|$ increases, verifying their robustness under versatile settings: instead of real-world Clean-Clean ER datasets, we now consider the synthetic Dirty ER datasets, and instead of SVC, we train our models using Weka's default implementation of Logistic Regression \cite{hall2009weka}.

The characteristics of the datasets, which are widely used in the literature \cite{DBLP:journals/tkde/Christen12,DBLP:journals/pvldb/0001SGP16}, appear in Table \ref{tab:datasets}b. To extract a large block collection from every dataset, we apply Token Blocking. In all cases, the recall is almost perfect, but precision and F1 are extremely low.

We consider four methods: \textsf{BCl} and \textsf{CNP} with the features and the training set size specified in \cite{papadakis2014supervised} as well as \textsf{BLAST} and \textsf{RCNP} with the features in Tables \ref{tab:full_comp_c_based}a and \ref{tab:full_comp_c_based}d, resp., trained over 50 labelled instances (25 per class). In each dataset, we performed three repetitions per algorithm and 
considered the average performance.

\begin{figure*}[t]
\centering
\includegraphics[width=0.49\linewidth]{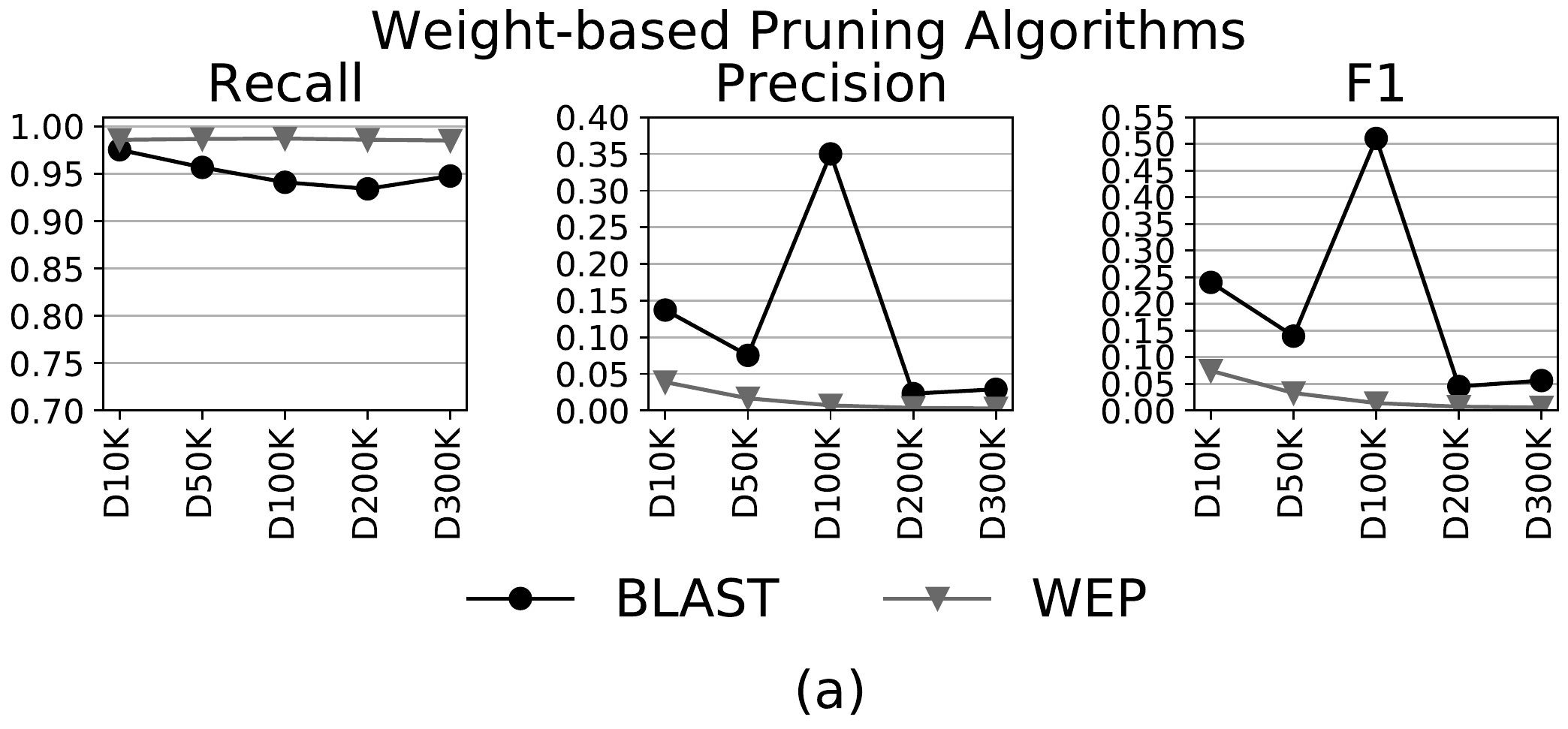}
\includegraphics[width=0.49\linewidth]{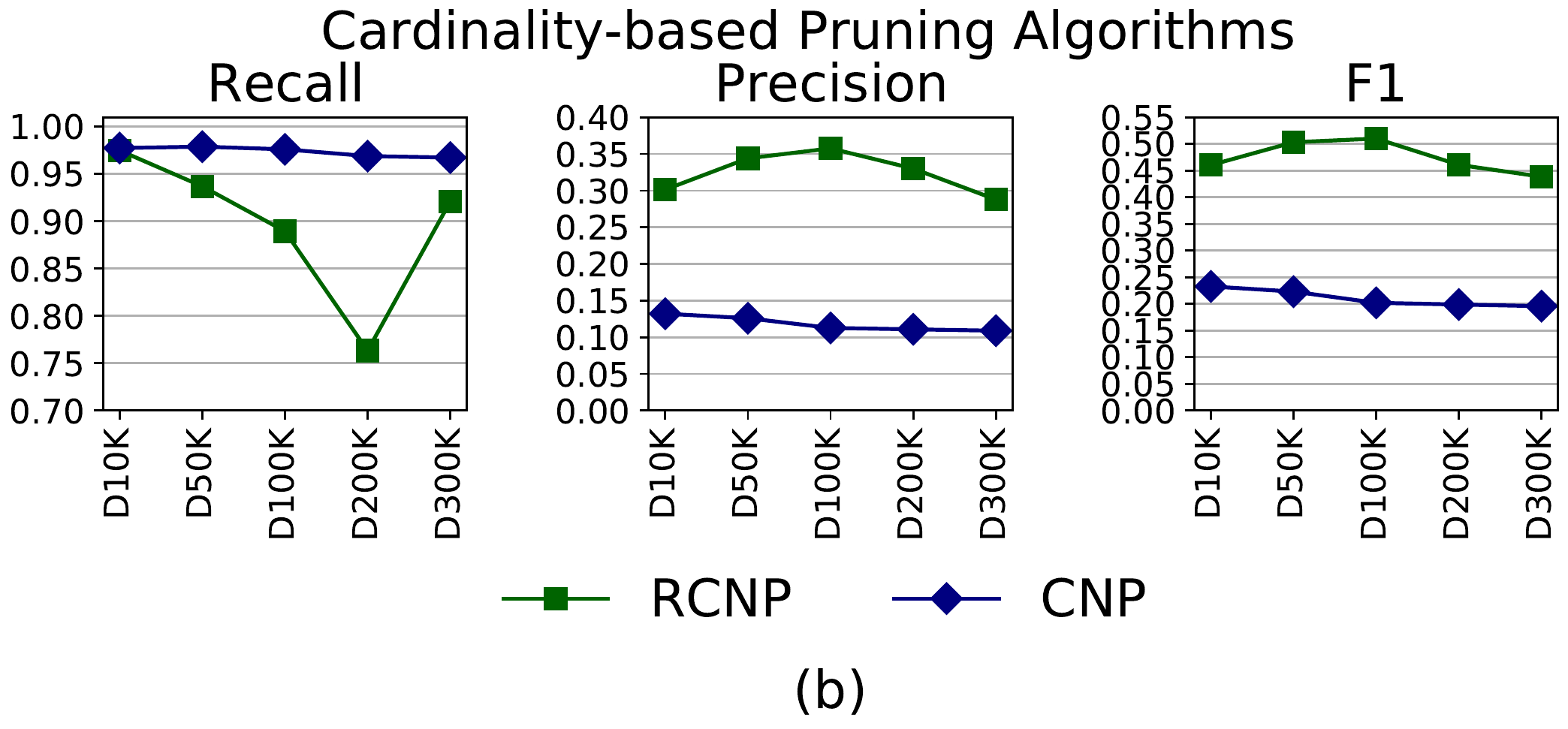}
\vspace{-10pt}
\caption{Scalability analysis over the datasets in Table \ref{tab:dirtyERdatasets}. (a) The effectiveness measures of \textsf{BCl}, which uses the features and the training set specified in \cite{papadakis2014supervised}, and \textsf{BLAST}, which combines the features in Formula \ref{feat:BLAST} with 50 randomly selected labelled instances, equally split between the two classes. (b) The effectiveness measures of \textsf{CNP}, which uses the features and the training set specified in \cite{papadakis2014supervised}, and \textsf{RCNP}, which combines the features in Formula \ref{feat:RCNP} with 50 randomly selected labelled instances, equally split between the two classes.}
\vspace{-15pt}
\label{fig:scalability}
\end{figure*}

The effectiveness of the weight- and cardinality-based algorithms over all datasets appear in Figures \ref{fig:scalability}a and \ref{fig:scalability}b, respectively. \textsf{BLAST} significantly outperforms \textsf{BCl} in all cases: on average, it reduces recall by 3.5\%, but consistently maintains it above 0.93, while increasing precision and F1 by a whole order of magnitude. 

{\color{\hcolor}
Note that the precision and F1 of BLAST is significantly higher than expected over $D_{100k}$, and vice versa for the recall of RCNP over $D_{200K}$.
Both cases should be attributed to the seed that is used for the random sampling of labelled pairs. 

In more detail, three iterations are performed per algorithm and dataset, with each one using a random set of 25 positive and 25 negative instances.
As a result, a different model is learned by logistic regression in each iteration.
Table \ref{tb:blastLR} shows the models corresponding to BLAST in the case of $D_{100K}$.
We observe significant variations in the factor of every feature across the three iterations.
Inevitably, there is a significant difference in the effectiveness of every model, which is determined by the number of candidate pairs and of detected duplicates. We actually observe that the first iteration produces a very high number of candidates, unlike the other two; the less candidates are produced, the less duplicates are detected. However, there is a smaller variation in recall than precision: the former fluctuates between 0.90 and 0.97 and the latter between 0.01 and 0.44. This explains the much higher precision (and F1) over $D_{100K}$, while recall remains in line with the rest of the datasets. It also indicates that Supervised BLAST assigns high probabilities to matching pairs regardless of the training set, thus being robust with respect to recall, which is very crucial for weight-based pruning algorithms. Note that different seeds yield a performance close to Iteration 1, which is the expected one, but we kept the seeds of Iteration 2 and 3 so as to show the effect of randomness.

\begin{table}[t]\centering
	\begin{tabular}{ | l | r | r | r |}
	    \cline{2-4}
	    \multicolumn{1}{c|}{} & Iteration 1 & Iteration 2 & Iteration 3\\
        \hline
        \hline
	    CF-IBF & -0.1814 & -0.1173 & -0.1522 \\
		RACCB & 10.8719 & -24.9254 & -7.3026 \\
		NRS	 & -1.3549 & -15.7979 & -1.7269 \\
		RS & -45.1 & -56.5986 & -130.687 \\
		Intercept & 41.7934 & 45.4136 & 56.702 \\
		\hline
		\hline
		Candidate pairs & 8,195,251	& 172,824 & 185,034 \\
		Detected duplicates & 83,031 & 76,868 & 79,263 \\
		\hline
	\end{tabular}
	\caption{BLAST's logistic regression models over $D_{100K}$.}
	\label{tb:blastLR}
\end{table}

Similar variations apply to the logistic regression models that lie at the core of Supervised RCNP. In this case, though, a specific number of the top-weighted candidates is retained per entity. A pair of entities $\langle e_i, e_j \rangle$ is then considered as a candidate pair if $e_j$ is among the top-weighted candidates of $e_i$ and vice versa. For this reason, different models might yield different sets of candidate pairs at the cost of lower recall, as in the case of $D_{200K}$.
}

\textsf{RCNP} outperforms \textsf{CNP} to a significant extent: on average, it reduces recall by 7.9\%, but maintains it to very high levels -- except for $D_{200K}$, where it drops to 0.77, {\color{\hcolor}due to the effect of random sampling}; yet, precision raises by 2.8 times and F1 by 2.3 times. These results verify the strength of our approaches, even though they require orders of magnitude less labelled instances than \cite{papadakis2014supervised}.

Most importantly, our approaches scale better to large datasets, as demonstrated by speedup in Figure \ref{fig:scalability}c.
Given two sets of candidate pairs, $|C_1|$ and $|C_2|$, such that $|C_1| < |C_2|$, this measure is defined as follows: $speedup = |C_2|/|C_1|\times RT_1 / RT_2$, where $RT_1$ ($RT_2$) denotes the running time over $|C_1|$ ($|C_2|$) -- in our case, $C_1$ corresponds to $D_{10K}$ and $C_2$ to all other datasets. In essence, speedup extrapolates the running time of the smallest dataset to the largest one, with values close to 1 indicating linear scalability, which is the ideal case. We observe that all methods start from very high values, but \textsf{BCl} and \textsf{CNP} deteriorate to a significantly larger extent than \textsf{BLAST} and \textsf{RCNP}, respectively, achieving the lowest values for $D_{300K}$. This should be attributed to their lower accuracy in pruning the non-matching comparisons, which deteriorates as the number of candidate pairs increases. As a result, they end up retaining and processing a much larger number of comparisons, which slows down their functionality.

{\color{\hcolor}
\textit{Overall, Generalized Supervised Meta-blocking scales much better to large datasets than Supervised Meta-blocking \cite{papadakis2014supervised} for both weight- and cardinality-based algorithms. For a slightly lower recall, it increases precision and F1 by at least 2 times, while maintaining a significantly higher speedup.}
}
\section{Related Work}
\label{sec:relatedWork}

\begin{table*}[t]
	\footnotesize
	\centering
	\begin{tabular}{lccccccccc}
		\cline{2-10}
		\multicolumn{1}{l|}{}                      & \multicolumn{1}{c|}{\textbf{AbtBuy}} & \multicolumn{1}{c|}{\textbf{DblpAcm}} & \multicolumn{1}{c|}{\textbf{ScholarDblp}} & \multicolumn{1}{c|}{\textbf{AmazonGP}} & \multicolumn{1}{c|}{\textbf{ImdbTmdb}} & \multicolumn{1}{c|}{\textbf{ImdbTvdb}} & \multicolumn{1}{c|}{\textbf{TmdbTvdb}} & \multicolumn{1}{c|}{\textbf{Movies}} & \multicolumn{1}{c|}{\textbf{WalmartAmazon}} \\ 
		\hline\hline
		\multicolumn{1}{|l|}{$Re$}      & \multicolumn{1}{c|}{0.8405}          & \multicolumn{1}{c|}{0.9759}           & \multicolumn{1}{c|}{0.9623}               & \multicolumn{1}{c|}{0.7358}            & \multicolumn{1}{c|}{0.8395}            & \multicolumn{1}{c|}{0.7465}            & \multicolumn{1}{c|}{0.8696}            & \multicolumn{1}{c|}{0.9275}          & \multicolumn{1}{c|}{0.9122}                 \\
		\multicolumn{1}{|l|}{$Pr$}   & \multicolumn{1}{c|}{0.1764}          & \multicolumn{1}{c|}{0.6463}           & \multicolumn{1}{c|}{0.3591}               & \multicolumn{1}{c|}{0.1264}            & \multicolumn{1}{c|}{0.3540}            & \multicolumn{1}{c|}{0.2325}            & \multicolumn{1}{c|}{0.1848}            & \multicolumn{1}{c|}{0.0992}          & \multicolumn{1}{c|}{0.0050}                 \\
		\multicolumn{1}{|l|}{$F1$}          & \multicolumn{1}{c|}{0.2914}          & \multicolumn{1}{c|}{0.7747}           & \multicolumn{1}{c|}{0.5190}               & \multicolumn{1}{c|}{0.2148}            & \multicolumn{1}{c|}{0.4971}            & \multicolumn{1}{c|}{0.3498}            & \multicolumn{1}{c|}{0.2954}            & \multicolumn{1}{c|}{0.1758}          & \multicolumn{1}{c|}{0.0100}                 \\
		\multicolumn{1}{|l|}{{$RT$ (sec)}} & \multicolumn{1}{c|}{6.20}            & \multicolumn{1}{c|}{5.67}             & \multicolumn{1}{c|}{11.73}                & \multicolumn{1}{c|}{6.83}              & \multicolumn{1}{c|}{6.55}              & \multicolumn{1}{c|}{6.77}              & \multicolumn{1}{c|}{8.32}              & \multicolumn{1}{c|}{126.13}          & \multicolumn{1}{c|}{107.56}                 \\ \hline
		\multicolumn{10}{c}{\textbf{(a) RCNP in combination with 50 balanced labelled instances and $\{CF$-$IBF, RACCB, JS, LCP, WJS\}$}}                                                                                                                                                                                                                                                                                                                                                                                         \\ \hline
		\multicolumn{1}{|l|}{$Re$}      & \multicolumn{1}{c|}{\color{\hcolor}0.8294}          & \multicolumn{1}{c|}{\color{\hcolor}0.9613}           & \multicolumn{1}{c|}{\color{\hcolor}0.9218}               & \multicolumn{1}{c|}{\color{\hcolor}0.7462}            & \multicolumn{1}{c|}{\color{\hcolor}0.8045}            & \multicolumn{1}{c|}{\color{\hcolor}0.7615}            & \multicolumn{1}{c|}{\color{\hcolor}0.8641}            & \multicolumn{1}{c|}{\color{\hcolor}0.8200}          & \multicolumn{1}{c|}{\color{\hcolor}0.7087}                 \\
		\multicolumn{1}{|l|}{$Pr$}   & \multicolumn{1}{c|}{\color{\hcolor}0.1797}          & \multicolumn{1}{c|}{\color{\hcolor}0.5984}           & \multicolumn{1}{c|}{\color{\hcolor}0.3745}               & \multicolumn{1}{c|}{\color{\hcolor}0.1031}            & \multicolumn{1}{c|}{\color{\hcolor}0.5471}            & \multicolumn{1}{c|}{\color{\hcolor}0.1867}            & \multicolumn{1}{c|}{\color{\hcolor}0.1720}            & \multicolumn{1}{c|}{\color{\hcolor}0.0090}          & \multicolumn{1}{c|}{\color{\hcolor}0.0002}               \\
		\multicolumn{1}{|l|}{$F1$}          & \multicolumn{1}{c|}{\color{\hcolor}0.2939}          & \multicolumn{1}{c|}{\color{\hcolor}0.7355}           & \multicolumn{1}{c|}{\color{\hcolor}0.5095}               & \multicolumn{1}{c|}{\color{\hcolor}0.1748}            & \multicolumn{1}{c|}{\color{\hcolor}0.6394}            & \multicolumn{1}{c|}{\color{\hcolor}0.2847}            & \multicolumn{1}{c|}{\color{\hcolor}0.2487}            & \multicolumn{1}{c|}{\color{\hcolor}0.0177}          & \multicolumn{1}{c|}{\color{\hcolor}0.0004}                 \\
		\multicolumn{1}{|l|}{{$RT$ }} & \multicolumn{1}{c|}{\color{\hcolor}5.95}            & \multicolumn{1}{c|}{\color{\hcolor}5.80}             & \multicolumn{1}{c|}{\color{\hcolor}11.33}                & \multicolumn{1}{c|}{\color{\hcolor}6.40}              & \multicolumn{1}{c|}{\color{\hcolor}5.91}              & \multicolumn{1}{c|}{\color{\hcolor}6.19}              & \multicolumn{1}{c|}{\color{\hcolor}6.89}              & \multicolumn{1}{c|}{\color{\hcolor}122.72}          & \multicolumn{1}{c|}{\color{\hcolor}107.62}                 \\ \hline
		\multicolumn{10}{c}{\color{\hcolor}\textbf{(e) CNP$_1$ with 50 labelled pairs and $\{CF$-$IBF, RACCB, JS, LCP, WJS\}$}}                                                                                                                                                                                                                                                                                                                                                                                  \\ \hline
		\multicolumn{1}{|l|}{$Re$}      & \multicolumn{1}{c|}{0.8347}          & \multicolumn{1}{c|}{0.9539}           & \multicolumn{1}{c|}{0.9581}               & \multicolumn{1}{c|}{0.7742}            & \multicolumn{1}{c|}{0.8345}            & \multicolumn{1}{c|}{0.7641}            & \multicolumn{1}{c|}{0.8677}            & \multicolumn{1}{c|}{0.9347}          & \multicolumn{1}{c|}{0.2332}                 \\
		\multicolumn{1}{|l|}{$Pr$}   & \multicolumn{1}{c|}{0.1895}          & \multicolumn{1}{c|}{0.6158}           & \multicolumn{1}{c|}{0.2184}               & \multicolumn{1}{c|}{0.0848}            & \multicolumn{1}{c|}{0.4132}            & \multicolumn{1}{c|}{0.1764}            & \multicolumn{1}{c|}{0.1484}            & \multicolumn{1}{c|}{0.0291}          & \multicolumn{1}{c|}{0.0001}               \\
		\multicolumn{1}{|l|}{$F1$}          & \multicolumn{1}{c|}{0.3081}          & \multicolumn{1}{c|}{0.7457}           & \multicolumn{1}{c|}{0.3453}               & \multicolumn{1}{c|}{0.1514}            & \multicolumn{1}{c|}{0.5247}            & \multicolumn{1}{c|}{0.2754}            & \multicolumn{1}{c|}{0.2363}            & \multicolumn{1}{c|}{0.0564}          & \multicolumn{1}{c|}{0.0002}                 \\
		\multicolumn{1}{|l|}{{$RT$ (sec)}} & \multicolumn{1}{c|}{15.61}           & \multicolumn{1}{c|}{9.64}             & \multicolumn{1}{c|}{28.51}                & \multicolumn{1}{c|}{13.63}             & \multicolumn{1}{c|}{11.37}             & \multicolumn{1}{c|}{9.99}              & \multicolumn{1}{c|}{11.41}             & \multicolumn{1}{c|}{1351.54}         & \multicolumn{1}{c|}{365.03}                 \\ \hline
		\multicolumn{10}{c}{\textbf{(c) CNP$_2$ in combination with the training set and the features proposed in \cite{papadakis2014supervised}, i.e., $\{CF$-$IBF, RACCB, JS, LCP\}$}}
	\end{tabular}
	\caption{Performance of the main cardinality-based algorithms across all datasets. 
		$RT$ is the mean run-time over~10~repetitions.}
	\vspace{-20pt}
	\label{tab:full_comp_c_based}
\end{table*}

The unsupervised pruning algorithms \textsf{WEP}, \textsf{WNP}, \textsf{CEP}, and \textsf{CNP} were introduced in \cite{DBLP:journals/tkde/PapadakisKPN14}. \textsf{WNP} and \textsf{CNP} were redefined in \cite{papadakis2016scaling} so that they do not produce block collections with redundant comparisons. 
Unsupervised Reciprocal \textsf{WNP} and Reciprocal \textsf{CNP} were coined in \cite{papadakis2016scaling}, while unsupervised \textsf{BLAST} was proposed in \cite{DBLP:journals/pvldb/SimoniniBJ16}.

Over the years, more unsupervised pruning algorithms have been proposed in the literature. \cite{zhang2017pruning} proposes a variant of \textsf{CEP} that retains the top-weighted candidate pairs with a cumulative weight higher than a specific portion of the total sum of weights. Crafted for Semantic Web data, MinoanER \cite{DBLP:conf/edbt/Efthymiou0SC19} combines meta-blocking evidence from two complementary block collections: the blocks extracted from the names of entities and from the attribute values of their neighbors. BLAST2 \cite{DBLP:journals/jdiq/BeneventanoBGS20} leverages loose schema information in order to boost the performance of Meta-blocking's weighting schemes. Finally, a family of pruning algorithms that focuses on the comparison weights inside individual blocks is presented in \cite{DBLP:journals/kais/NascimentoPM20}; for example, Low Entity Co-occurrence Pruning removes from every block a specific portion of the entities with the lowest average weights. Our approaches can be generalized to these algorithms, too, but their analytical examination lies out of our scope.

The above works consider Meta-blocking in a static context that ignores the outcomes of Matching. A dynamic approach that leverages Meta-blocking to make the most of the feedback of Matching is \textit{pBlocking} \cite{DBLP:journals/vldb/GalhotraFSS21}. After applying Matching to the smallest blocks, intersections of the initial blocks are formed and scored based on their ratio of matching and non-matching entities. Meta-blocking is then applied to produce the next set of candidate pairs that will be processed by Matching. This process is iteratively applied until convergence. \textit{BEER} \cite{DBLP:conf/sigmod/GalhotraFSS21} is an open-source tool that implements pBlocking.

The work closest to ours is BLOSS \cite{DBLP:journals/is/BiancoGD18}. It introduces  an active learning approach that reduces significantly the size of the labelled set required by Supervised Meta-blocking. Initially, it partitions the unlabelled candidate pairs into similarity levels based on \textsf{CF-IBF}. Then, it applies rule-based active sampling inside every level in order to select the unlabelled pairs with the lowest commonalities with the already labelled ones so as to maximize the captured information. In the final step, BLOSS cleans the labelled sample from non-matching outliers with high Jaccard weight. 


\begin{figure}[t]
\centering
\includegraphics[width=0.8\linewidth]{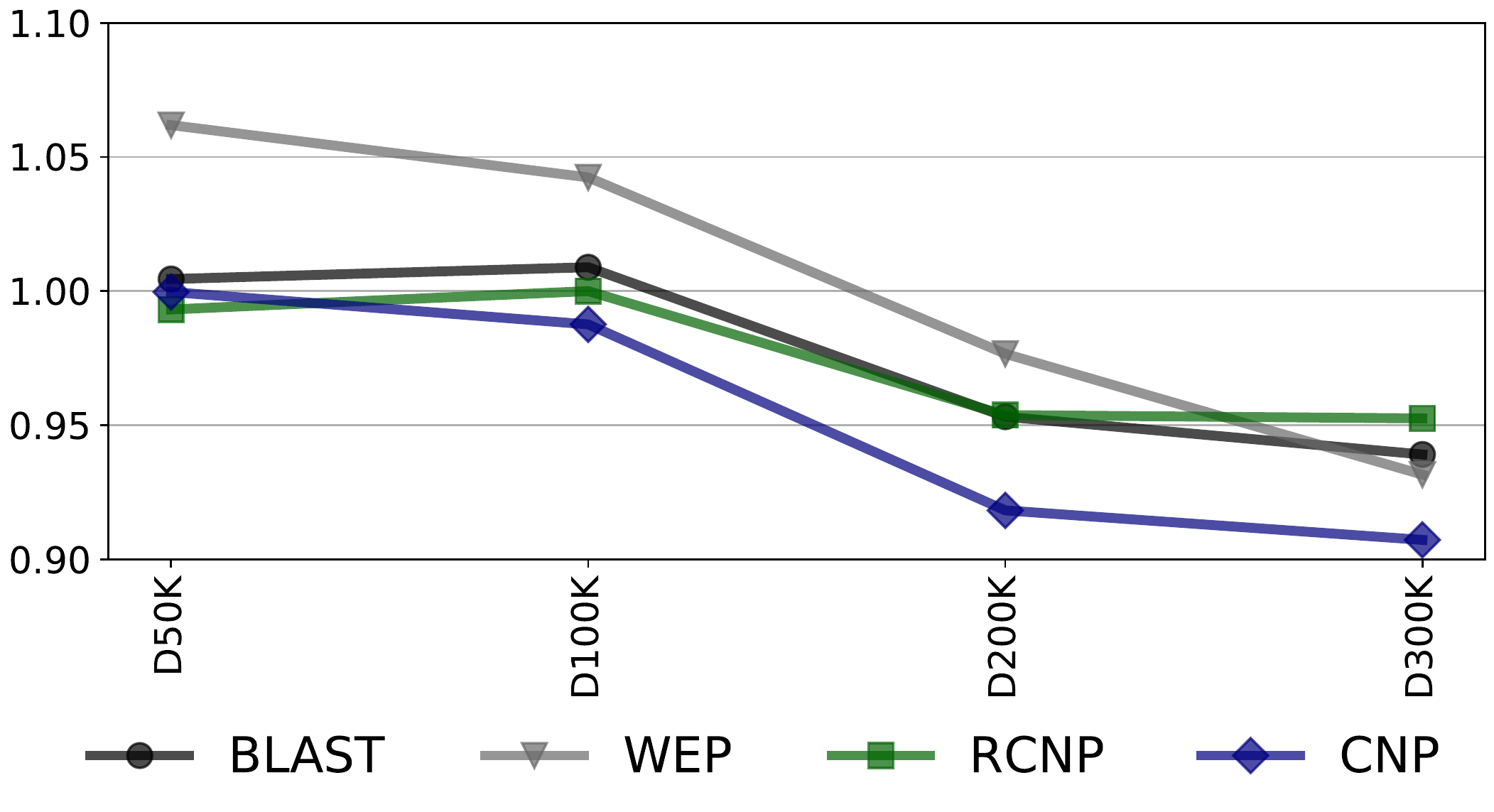}
\vspace{-5pt}
\caption{The speedup of the algorithms used in the scalability analysis of Figure \ref{fig:scalability}.}
\vspace{-15pt}
\label{fig:speedup}
\end{figure}

\section{Conclusions}
\label{sec:conclusions}

We have presented Generalized Supervised Meta-blocking, which casts Meta-blocking as a probabilistic binary classification task and weights all candidate pairs in a block collection with the probabilities produced by the trained classifier. These weights are processed by a pruning algorithm that can be: (i) weight-based, determining the minimum weight of retained pairs in a way that promotes recall, or (ii) cardinality-based, determining the maximum number of retained pairs in a way that promotes precision. Through a thorough experimental study over 9 established, real-world datasets, we verified that \textsf{BLAST} and \textsf{RCNP} constitute the best weight- and cardinality-based pruning algorithms, respectively. 
We also demonstrated that four new weighting schemes give rise to feature sets that outperform the one determined in~\cite{papadakis2014supervised} as optimal. Finally, we showed that a very small, balanced training set with just 50 labelled instances suffices for consistently achieving high effectiveness, high time efficiency and high scalability.

In the future, we plan to leverage Generalized Supervised Meta-blocking as a means for optimizing the performance of Progressive Entity Resolution~\cite{DBLP:journals/tkde/SimoniniPPB19}.


\bibliographystyle{ACM-Reference-Format}
\bibliography{references}

\end{document}